\documentclass[useAMS,usenatbib]{mn2e}

\usepackage{epsfig}

\newcommand{\msun}{\mbox{$M_\odot$}}

\title[Star Formation Histories in Barred Spiral Galaxies]
{The Mass Dependence of Star Formation Histories 
in Barred Spiral Galaxies}
\author[Christian Carles et al.]
{Christian Carles,$^{1,2}$ Hugo Martel,$^{1,2}$ Sara L. Ellison,$^3$
and Daisuke Kawata$^4$\\
$^{1}$D\'epartement de physique, de g\'enie physique et d'optique,
Universit\'e Laval, Qu\'ebec, QC, G1V 0A6, Canada\\
$^{2}$Centre de Recherche en Astrophysique du Qu\'ebec,
C.P. 6128, Succ. Centre-Ville, Montr\'eal, QC, Canada\\
$^3$Department of Physics and Astronomy, University of Victoria,
Victoria, BC, Canada\\
$^4$Mullard Space Science Laboratory, University College London,
Holmbury St. Mary, Dorking, Surrey, UK}
\begin{document}

\date{Accepted XXX. Received XXX; in original form XXX}

\pagerange{\pageref{firstpage}--\pageref{lastpage}} \pubyear{XXX}

\maketitle

\label{firstpage}

\begin{abstract}
We performed a series of 29 gasdynamical simulations of disc galaxies, 
barred and unbarred, with various stellar masses, to
study the impact of the bar on star formation history. Unbarred galaxies evolve 
very smoothly, with a star formation rate (SFR)
that varies by at most a factor
of three over a period of $2\,\rm Gyr$. The evolution of barred galaxies
is much more irregular, especially at high stellar masses.
In these galaxies, the bar drives a substantial
amount of gas toward the centre, resulting in a high SFR,
and producing a starburst in the most massive galaxies. 
Most of the gas
is converted into stars, and gas exhaustion leads to a rapid drop of
star formation after the starburst. 
In massive barred galaxies 
(stellar mass $M_*>2\times10^{10}\msun$) the large amount
of gas funnelled toward the centre is completely consumed by the starburst,
while in lower-mass barred galaxies it is only partially consumed. 
Gas concentration is thus higher in lower-mass barred galaxies 
than it is in higher-mass ones.
Even though unbarred galaxies funnelled less gas toward their centre,
the lower SFR allows this gas to accumulate.
At late times, the star formation efficiency is higher in barred galaxies 
than unbarred ones, enabling these galaxies to maintain a higher SFR with 
a smaller gas supply.
Several properties, such as the global 
SFR, central SFR, or central gas concentration, vary monotonically
with time for unbarred galaxies, but not for barred galaxies. 
Therefore one must be careful
when comparing barred and unbarred galaxies that share one observational
property, since these galaxies might be at very different stages 
of their respective evolution.
\end{abstract}


\section{INTRODUCTION}

Bars are one of the most prominent morphological and evolutionary structures
of spiral galaxies. Their presence is quite ubiquitous, with recent
observational data consistently establishing a bar fraction somewhere
between 30\% and 60\% in the local universe
\citep{knapen_subarcsecond_2000,martinez_relating_2011,masters_galaxy_2011,
lee_dependence_2012} and a slightly lesser value at higher redshift
\citep{elmegreen_constant_2004,sheth_evolution_2008,
simmons_galaxy_2014}. Bars create a strong torque on the galaxy
\citep{lynden-bell_mechanism_1979,athanassoula_angular_2003} leading to a
redistribution of the gaseous and stellar component
\citep{gadotti_homogenization_2001,grand_impact_2015} and a transport of
angular momentum from the inner to the outer regions of the galaxy
\citep{debattista_constraints_2000,athanassoula_what_2003,
martinez-valpuesta_evolution_2006,kim_two-dimensional_2012,
lokas_adventures_2014,seidel_balrog_2015}. 
These dynamical effects have a wide range of consequences
on both the gaseous and stellar content of the galaxy: the central
bulge-like stellar component gets heated up \citep{berentzen_gas-driven_1998,
fathi_bulges_2003,kormendy_secular_2004,
athanassoula_nature_2005,berentzen_gas_2007}, 
while gas flows from the outer to the inner regions of the galaxy 
\citep{combes_spiral_1985,combes_bars_1993,
maciejewski_gas_2002,regan_bar-driven_2004,
baba_interpretation_2010,masters_galaxy_2012,kubryk_evolution_2015}.
As a result, the radial chemical abundance profile initially
flattens 
\citep{vila-costas_relation_1992,martin_influence_1994,
di_matteo_signatures_2013},
and the concentration of gas in the centre of the
galaxy increases \citep{knapen_central_1995,sakamoto_bar-driven_1999}. 
The accumulation of gas in the centre eventually triggers an increase
in the star formation rate (SFR) \citep{devereux_stellar_1987,
martin_quantitative_1995,martinet_bar_1997,alonso-herrero_statistical_2001,
hunt_molecular_2008,coelho_bars_2011}. Simulations have shown 
that this enhanced star formation activity
increases the metallicity of the gas in the central region
\citep{friedli_influence_1994,friedli_secular_1995},
thus explaining the origin of the observed
break in the slope of the chemical abundance profiles of
barred galaxies
\citep{martin_oxygen_1995,roy_abundance_1997,considere_starbursts_2000}.

These studies reveal the importance of the bar in the
secular evolution of the host galaxy. 
Although the general overview of the impact of a bar on a galactic disc is
well accepted, the details of the process, and its dependence on galactic 
properties, pose many remaining questions.
While the
presence of a bar often causes an increase of the SFR, several observations
show no increase \citep{pompea_test_1990,
martinet_bar_1997,chapelon_starbursts_1999} or an increase only in
early-type spiral galaxies
\citep{ho_influence_1997,james_h_2009}. Recent work
with the volunteer-based identification of morphological types by the
Galaxy Zoo team showed that the specific
star formation rate (SSFR) is
anti-correlated with the presence of a bar \citep{cheung_galaxy_2013}
but at same time the relation between the SFR
and the stellar mass $M_*$ seems unaffected by the
presence of a bar
\citep{willett_galaxy_2015}. This matter is further complicated
by bar strength and length: Early-type spirals tend to have stronger
bars \citep{elmegreen_relative_1989,erwin_how_2005,
menendez-delmestre_near-infrared_2007}, longer bars are redder and
situated in redder galaxies \citep{hoyle_galaxy_2011}, and bar length is
correlated with stellar mass, S\'ersic index, and central surface star
density \citep{cheung_galaxy_2013}, all of which affect the SFR,
though some studies found only marginally higher star formation
efficiencies in galaxies with strong bars \citep{saintongeetal12}. The SFR of
barred galaxies has also been shown to correlate with the central gas
mass \citep{jogee_central_2005},  and the increase in SFR, or lack thereof,
in barred galaxies could be due to a temporarily pre- or
post-starburst phase \citep{martinet_bar_1997}.
A very recent study by \citet{sandstrometal16} shows that higher SFRs
in the central kpc of barred galaxies are not caused by larger central gas 
supplies, but instead by much higher star formation efficiencies compared
to unbarred galaxies, in contradiction with earlier results by
\citet{sakamoto_bar-driven_1999}.

Metallicities and gas content  
have also yielded conflicting results,
with wide
variations in radial abundance profiles in both gas 
\citep{edmunds_co-existence_1993,oey_abundances_1993,zaritsky_h_1994,
considere_starbursts_2000} and stars \citep{perez_study_2009,
sanchez-blazquez_star_2011,seidel_balrog_2016}.
 Comparison between barred and unbarred
galaxies show that the former can have higher central metallicities 
\citep{ellison_impact_2011}, no significant variations
\citep{cacho_gaseous-phase_2014,sanchez-blazquezetal14}, 
or even lower metallicities than the
latter \citep{dutil_chemical_1999,considere_starbursts_2000}.
\citet{masters_galaxy_2012} showed that bars are more common in gas-poor
galaxies, but it is unclear if this is due to SFR-related gas exhaustion
triggered by the bar, by lower bar formation in gas-rich galaxies or if
both gas fraction and bar formation are correlated to environmental effects 
\citep{martinez-valpuesta_characterization_2016}.

The observational results described above indicate that the effect of a 
bar on the re-distribution of gas and metals, and subsequent star formation, 
is a highly complex process that proceeds differently in different 
galaxies. The mass
of the host galaxy, its morphological type, its gas fraction, the strength
and length of the bar, and the presence of an AGN are all important
factors which can potentially impact the local and global star formation
history, the radial migration of stars, the gas enrichment and its flows.
In previous work (\citealt{martel_connection_2013}, hereafter Paper~I)
we simulated a barred galaxy with a mass comparable to that of the Milky Way,
and showed that the presence of
the bar causes an important enhancement of the metallicity of gas situated
within the {\it central region\/}, 
defined as a sphere with radius of $1\,\rm kpc$.
We also showed that this enhancement
could not be solely attributed to the local star formation in the central
region, but that 50\% of these metals originated from a different
location and flowed toward the centre along the bar.

In the current paper, we extend the work of Paper~I by simulating a suite of 
barred and unbarred galaxies with a range of stellar masses and gas fractions.
This work is particularly motivated by the observational result of 
\citet{ellison_impact_2011}
that the SFRs of barred galaxies are enhanced only for galaxies with 
$\log(M_*/\msun)>10$, where $M_*$ is the stellar mass. 
In contrast, the metallicity enhancement measured by 
\citet{ellison_impact_2011} is seen at all stellar masses.
The disparity between the central metallicity and the
SFR was believed to be due to a fast, starburst-like formation episode
in low-mass galaxies while high-mass galaxies retained a high SFR through
their evolution. 
In order to provide a theoretical comparison with the results of 
\citet{ellison_impact_2011}, we simulate isolated
barred and unbarred galaxies of various stellar masses between
$\log M_*=9.6$ and 10.4  and study the variation as a function of mass of
the SFR and gas flows along the bar of both barred and unbarred galaxies.
We present our simulation code and
our suite of simulations in Section~\ref{sec:CGCP}. Results are presented
in Section~\ref{sec:results}. Summary and conclusion are presented
in Section~\ref{sec:conc}. 

\section{THE SIMULATIONS}
\label{sec:CGCP}

\subsection{The numerical algorithm}
\label{sec:num}

All the simulations in this paper were performed using the numerical
algorithm GCD+ \citep{kawata_gcd+:_2003,rahimi_towards_2012,
kawata_calibrating_2013,kawata_numerical_2014}. GCD+ is a three-dimensional
tree/smoothed particle hydrodynamics (SPH)
algorithm \citep{lucy_numerical_1977,gingold_smoothed_1977} which
simulates the galactic chemodynamical evolution, accounting for
hydrodynamics, self-gravitation, star formation, supernovae feedback,
metal enrichment and diffusion, and radiative cooling. It uses an
artificial thermal conductivity suggested by \citet{rp07} 
to resolve the Kelvin-Helmholtz instability,
and the adaptive softening length suggested by
\citet{price_energy-conserving_2007}.
Metal diffusion is computed using the method of \citet{greif_chemical_2009},
while radiative cooling and heating are handled using tables computed with
CLOUDY \citep{ferland_cloudy_1998,robertson_molecular_2008}. Star formation
is handled by transforming gas particles into star particles as described
in \citet{kawata_numerical_2014}: if the local velocity of the gas particle
is convergent and the density exceeds a given density threshold
$n_{\rm th}^{\phantom1}$, the gas particle may transform into a star particle
with a probability weighted by its density. The star particles are assumed 
to consist of stars whose mass follow a \citet{salpeter_luminosity_1955}
initial mass function and the metal enrichment they produce from 
Type~II and Ia supernovae is calculated from
\citet{woosley_evolution_1995} and 
\citet{iwamoto_nucleosynthesis_1999}.

Four main parameters govern the star formation rate and the supernovae
feedback \citep{rahimi_towards_2012} and are fixed as follows: the
supernovae energy output $E_{\mathrm{SN}}=1.439\times10^{50}\mathrm{erg}$,
the stellar wind energy output 
$E_{\mathrm{SW}}=5.0\times 10^{36}\rm erg\,s^{-1}$,
the star formation efficiency $C_*=0.02$, and the star formation density
threshold $n_{\rm th}^{\phantom1}=1\,\rm cm^{-3}$.

\subsection{Initial conditions}

For generating the initial conditions of our simulations, we use the same
technique as in \citet{grand_impact_2015}. The system consists of a dark 
matter halo which is treated analytically, and a disc made of gas and stars,
which are represented by stellar and gaseous particles. We do not include a
central bulge in the initial conditions.
We set up the stellar particle
disc using an exponential surface density profile:
\begin{equation}
\rho_* = \frac{M_*}{4 \pi z_* R^2_*}{\rm sech}^2
\left(\frac{z}{z_*}\right)\exp\left(-\frac{R}{R_*}\right),
\label{eq:Stellardisk}
\end{equation}

\noindent
where $M_*$ is the stellar disc mass, $R_*$ its scale length, and $z_*$ the
scale height, and $R$ and $z$ are the radial and vertical coordinates,
respectively. The gaseous disc has the same radial exponential surface
density, but its height is determined by imposing an initial hydrostatic
equilibrium within the gaseous disc. We then set an initial radial
metallicity profile in both the stellar and gaseous populations, with the
iron abundance being given by 
\begin{equation}
\label{eq:FeHprofile}
\mathrm{[Fe/H]}= 0.2 -0.05R,
\end{equation}

\noindent
where $R$ is in kpc. $\alpha-$elements are initially only present in the
stellar component and their abundance is given by 
\begin{equation}
\label{eq:alphaFeprofile}
{\rm[\alpha/Fe]= -0.16[Fe/H]}(R).
\end{equation}

We modify the metallicity of each particle by adding a gaussian scatter of 
$0.02\,\rm dex$ to create a local dispersion of their abundances. The star
particles are assigned an initial age using an age-metallicity relation 
$\rm[Fe/H]=-0.04\times\,age(Gyr)$.
As in Paper~I, we do not simulate the evolution of the
dark matter halo. Instead, we assume a static halo with an NFW profile
\citep{nfw96}, which is appropriate for simulations of isolated galaxies.

As we vary the mass of the stellar component of the galaxy, we must adapt
all of the other parameters, such as the size of the stellar disc and the
mass and size of the gaseous and dark matter components, to follow the
expected behaviour. Following \citet{cox_feedback_2006},
we derive the values of all these parameters using observational or
simulated relations between one parameter and the other, thus obtaining
an ``average'' galaxy of a given stellar mass. We first derive the mass of
the NFW halo using the results from
\citet{moster_constraints_2010}, who obtained a parametrisation of 
the ratio between dark matter halo mass and the stellar mass within the
halo using abundance matching analysis. This parametrisation takes the form:
\begin{equation}
\frac{M_*}{M_{200}}=2\left(\frac{M_*}{M_{200}}\right)_0
\left[\left(\frac{M_{200}}{M_1}\right)^{-\beta}+
\left(\frac{M_{200}}{M_1}\right)^{\gamma}\right]^{-1},
\end{equation} 

\noindent
where $M_{200}$ is the halo mass, 
$(M_*/M_{200})_0$ is a normalisation factor, $M_1$ the transition mass
between an evolution as a power of $\beta$ and $\gamma$. We use 
$\log M_1=11.899$, $(M_*/M_{200})_0=0.002817$, $\beta=1.068$, and 
$\gamma=0.611$ as suggested by the best fit in
\citet{moster_constraints_2010}. Instead of attributing directly a
scale length to the NFW halo, we fix the concentration parameter
$c=r_{200}/r_s$ to 8 in barred galaxies and 20 in unbarred galaxies.
As shown in \citet{athanassoula_morphology_2002} and Paper~I, a high
concentration parameter stabilizes the simulated disc, preventing the
formation of the otherwise naturally-occurring bar.
 We use the halo mass and scale length
 to calculate a fixed gravitational potential which will
act on star and gas particles through the simulation. By forgoing the
dynamical nature of the dark matter halo we can greatly increase our
baryonic resolution for a given computational time while having little
impact on star formation and gas evolution.
The value of the concentration parameter, and the presence of
a live dark matter halo, have been shown 
to influence the rotation speed, size, morphology, and stability of the bar 
during its evolution \citep{athanassoula_morphology_2002,
holley-bockelmann_bar-induced_2005,martinez-valpuesta_evolution_2006,
sellwood_bar_2016}. However, their influence are prominent in the 
long-term evolution of the host galaxy (e.g. over a Hubble time), while 
we focus here on the shorter-term, initial effect of the bar on the star 
formation history, i.e. $2\,\rm Gyr$ after bar formation.
Long-term evolution will be considered in future work.

The scale radius of the stellar disc is calculated using the relation 
between $R_{50}$, the half-light radius, and $M_*$, the mass of the 
stellar disc, as found by \citet{shen_size_2003}: 
\begin{equation}
R_{50}(\mathrm{kpc})=\gamma M_*^{\alpha} 
\left(1+\frac{M_*}{M_0}\right)^{\beta - \alpha},
\end{equation}

\noindent
where $\gamma$ is a scaling factor, $M_0$ is the characteristic mass of the 
transition between the relation for lower-mass galaxies and that
for higher-mass galaxies. 
We use $\gamma=0.1$, $M_0=3.98\times10^{10}\msun$, 
$\alpha = 0.14$, and $\beta = 0.39$ to evaluate the half-light radius for a
given mass. Assuming that the half-light radius corresponds roughly with the 
half-mass radius, we integrate the density profile of the stellar disc 
(eq. \ref{eq:Stellardisk}) up 
to $R_{50}$ to obtain a transcendental relation between $R_*$ and $R_{50}$ 
which lets us compute the scale-radius from the disc mass.

Finally, we set the values of the parameters for the gaseous disc using the 
same relation as \citet{cox_feedback_2006},
\begin{equation}
\log M_{\rm gas}=0.78\log M_*-1.74 ,
\end{equation}

\noindent
where both masses are expressed in $10^{10}\msun$ units. Since this relation
has a non-zero initial value, the gas fraction of the galaxies will vary
with stellar mass, with low-mass galaxies having a higher gas fraction than
high-mass ones. As for the gaseous disc scale length, it is fixed at twice
the scale length of the corresponding stellar disc.

\subsection{Runs and parameters}
\label{sec:inc}

\begin{table*}
 \centering
 \begin{minipage}{140mm}
  \caption{Initial parameters of the simulations.
           All masses are in units of $10^9\msun$}
  \begin{tabular}{@{}crrrrrcl@{}}
  \hline
Galaxy & $M_*$ & $M_{\rm gas}$ & $M_{200}$ & 
$N_*$ & $N_{\rm gas}$ & $f_{\rm gas}$ & colour \\
  \hline
O &  4.0 &  1.72 &  265 &  52 501 &  22 546 & 0.300 & purple      \\
A &  5.0 &  2.04 &  299 &  66 584 &  27 244 & 0.289 & black       \\
B &  6.3 &  2.45 &  341 &  85 121 &  33 078 & 0.279 & blue        \\
C &  7.9 &  2.92 &  389 & 108 210 &  40 008 & 0.269 & turquoise   \\
D & 10.0 &  3.51 &  450 & 138 869 &  48 749 & 0.259 & light green \\
E & 12.5 &  4.18 &  519 & 175 774 &  58 748 & 0.250 & dark green  \\
F & 15.8 &  5.02 &  609 & 225 010 &  71 426 & 0.241 & yellow      \\
G & 20.0 &  6.03 &  726 & 288 333 &  86 901 & 0.231 & red         \\
H & 25.0 &  7.17 &  872 & 364 460 & 104 583 & 0.222 & burgundy    \\
I & 50.0 & 12.30 & 1848 & 752 656 & 185 430 & 0.197 & brown       \\
   \hline
\end{tabular}
\label{table:initial}
\end{minipage}
\end{table*}

We performed an initial series of 19 simulations of isolated galaxies,
evolving over a period of $\rm2\,Gyr$.
Since one of our goals is to understand the origin of the
$10^{10}\msun$ transition in the SFR described in \citet{ellison_impact_2011},
we sample masses both inferior and superior to this transition value, with
stellar masses going from $\log(M_*/\msun)=9.6$ to 10.4. Note that the
most massive barred galaxy has the same mass as the ones simulated in Paper~1.
We simulated 9 galaxies with stellar masses ranging from
$4\times10^9\msun$ to $25\times10^9\msun$. We name these galaxies
O-A-B-C-D-E-F-G-H. For each mass, we performed two simulations.
In the first simulation,
we used a dark matter halo with a concentration
parameter $c=8$ to allow a bar to form naturally due to instabilities
In the second simulation, we increased the concentration parameter
to $c=20$. This stabilizes the disc and prevents the formation of
a bar (see Paper~I). Hence, for each global stellar mass, we have two galaxies, 
a barred one and an unbarred one.
However, we also want to compare galaxies
with comparable {\it central\/} stellar
mass (the mass inside the $1\,\rm kpc$ 
central region). 
Because barred galaxies have a higher central mass density at fixed
$M_*$ than unbarred ones,
we performed one additional simulation
of an unbarred galaxy with a stellar mass
$50\times10^9\msun$, named galaxy I. This galaxy
achieves a similar central mass as runs G and H.
The number of particles in each simulation is chosen in order to have a
comparable mass resolution in all simulations while maintaining a reasonable
number of particles for both low-mass galaxies, as a minimal number of
particles are needed to have realistic results, and high-mass galaxies,
where computing time limits our maximal number of particles. 


The values of the parameters are shown
in Table~\ref{table:initial}. 
$M_*$, $M_{\rm gas}$, and $M_{200}$ are the initial stellar mass, initial
gas mass, and virial mass of the galaxy, respectively, $N_*$ and $N_{\rm gas}$
are the initial number of star and gas particles, respectively, and 
$f_{\rm gas}\equiv M_{\rm gas}/(M_*+M_{\rm gas})$ is the initial gas fraction.
The last column 
refers to the colours used throughout the figures of this paper to
distinguish the various runs.
We use a heat colour-coding where bluer, colder 
colours represent low-stellar mass galaxies and red, hotter colours represent 
more massive ones.
Notice that $M_*$, $M_{\rm gas}$, and $f_{\rm gas}$ evolve with time as gas in
converted into stars. The values in Table~\ref{table:initial} 
are the initial values,
which are the ones appearing in equations~(1)--(6). In the
remainder of the paper, the symbols $M_*$, $M_{\rm gas}$, and $f_{\rm gas}$
refer to the values at the epoch of interest.

\section{RESULTS}
\label{sec:results}

\subsection{Global properties}
  
\begin{figure}
\centering
\includegraphics[scale=0.22]{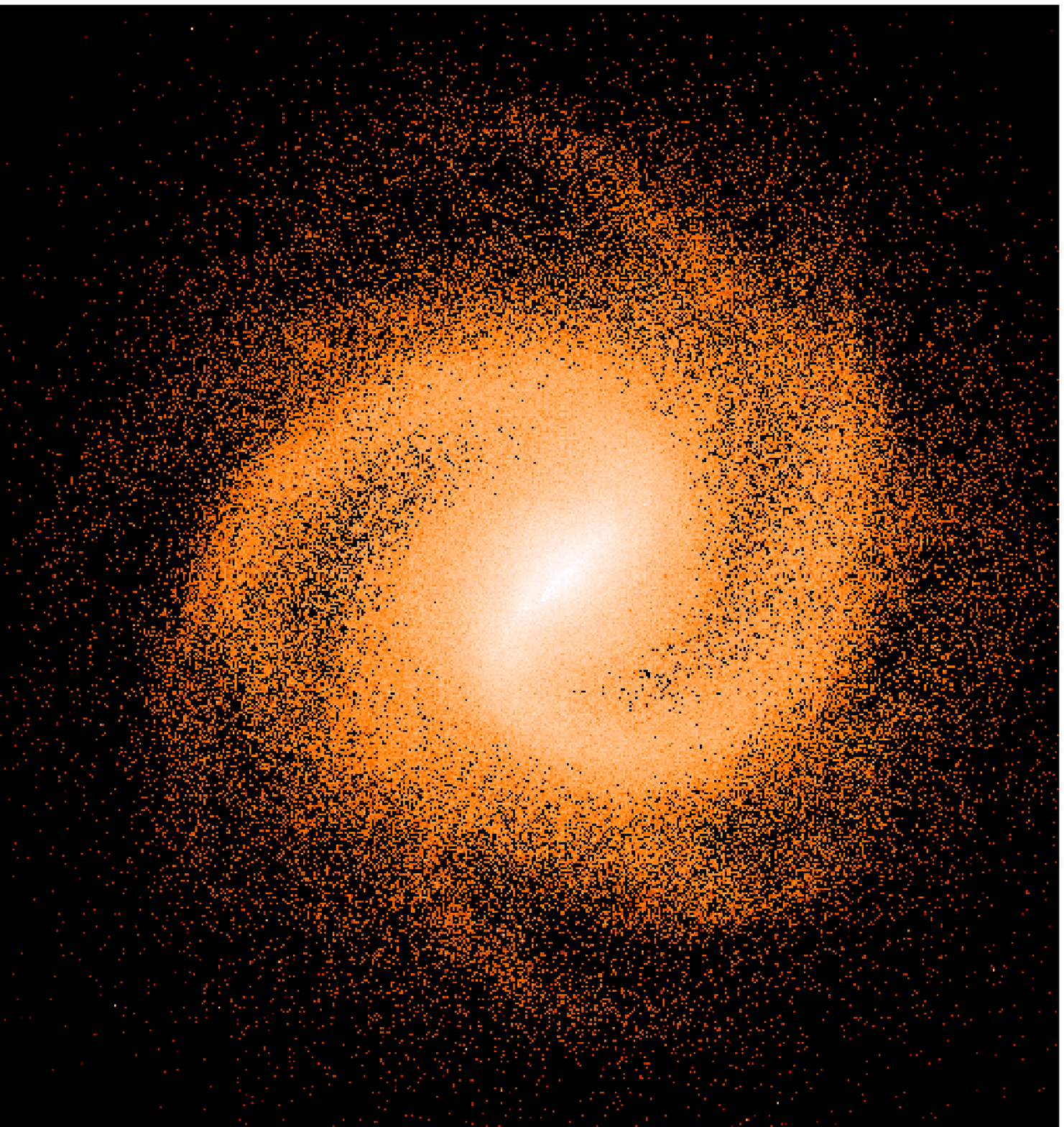}
\includegraphics[scale=0.22]{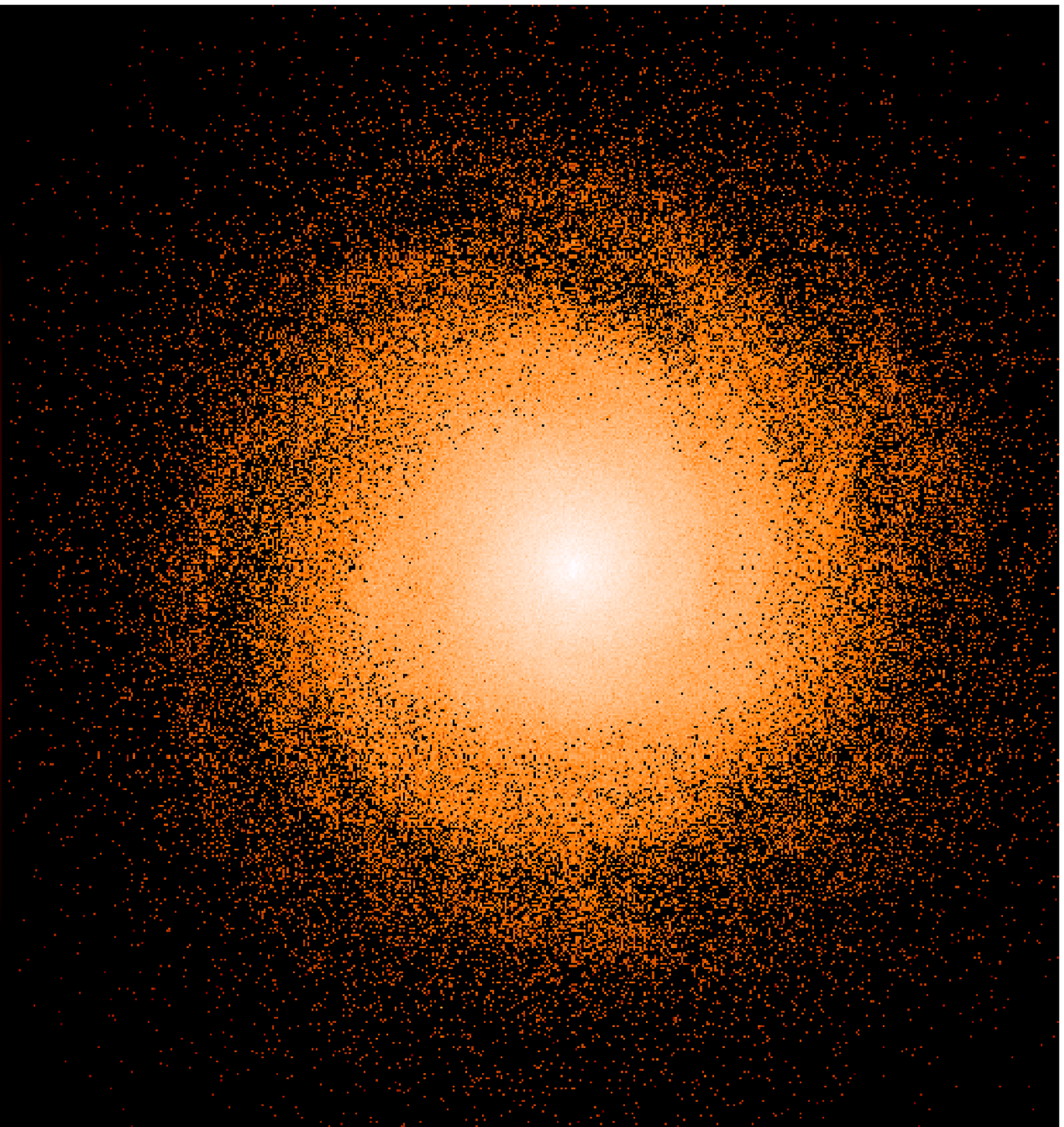}
\vskip2pt
\includegraphics[scale=0.22]{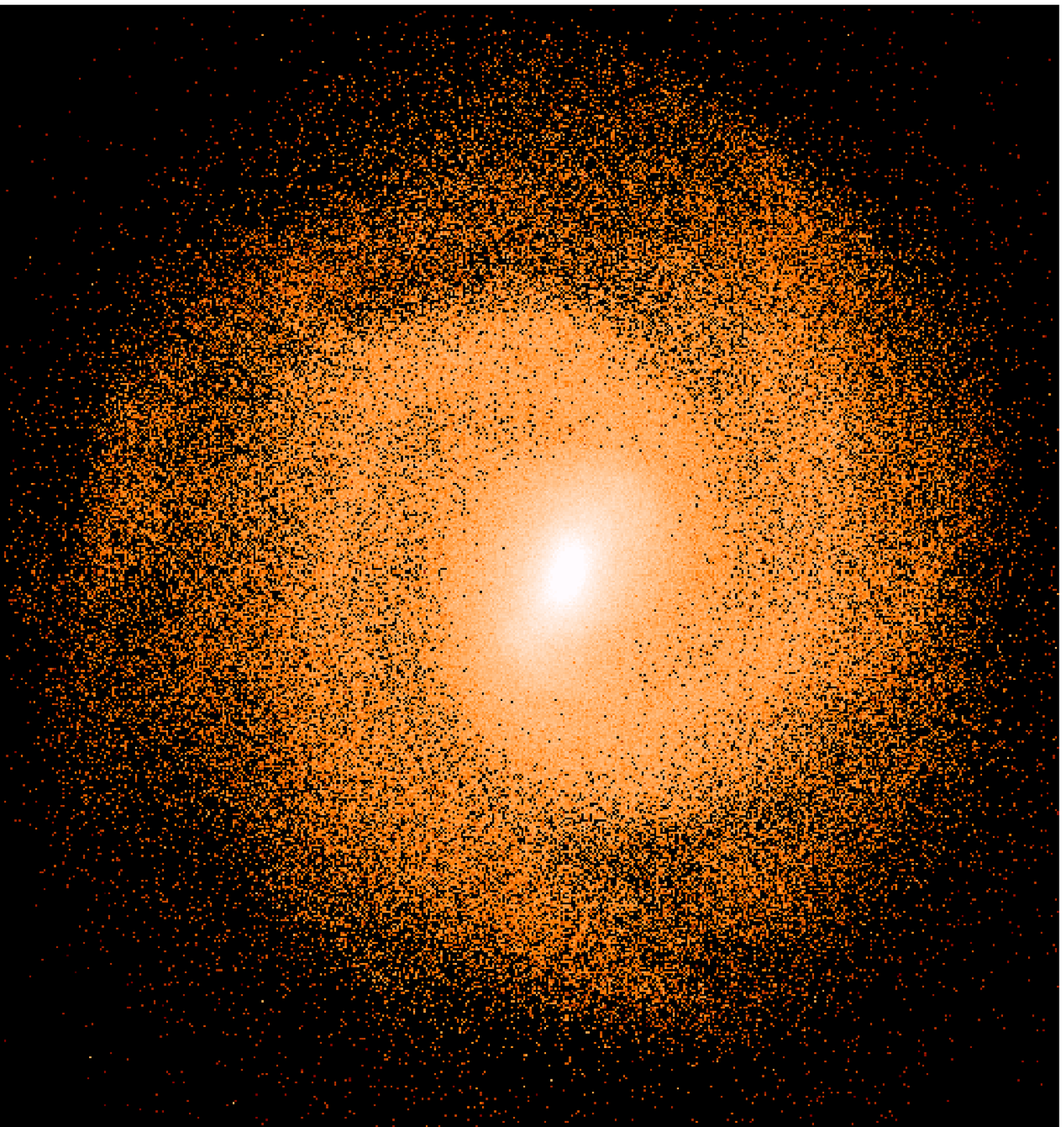}
\includegraphics[scale=0.22]{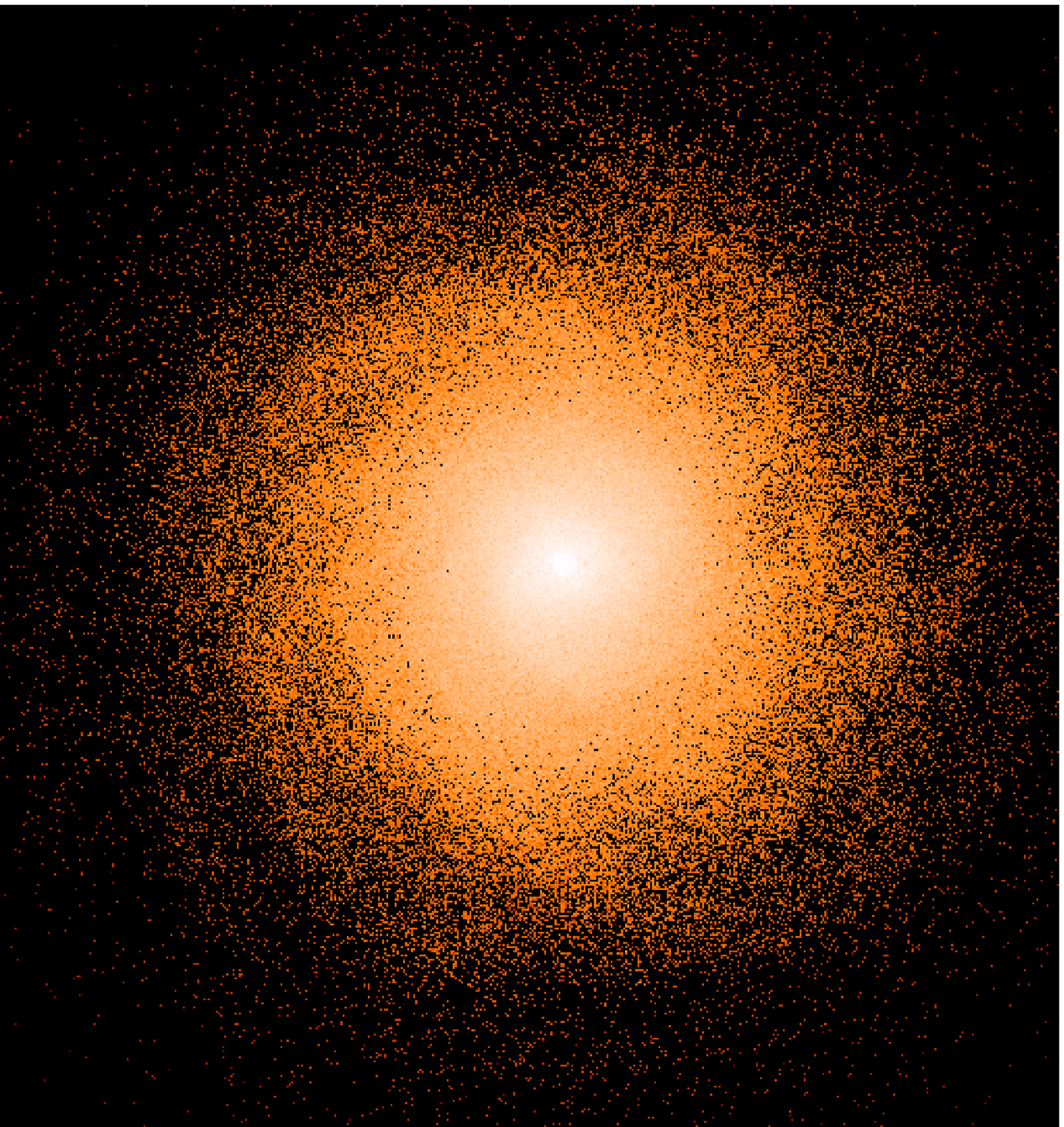}
\vskip2pt
\includegraphics[scale=0.22]{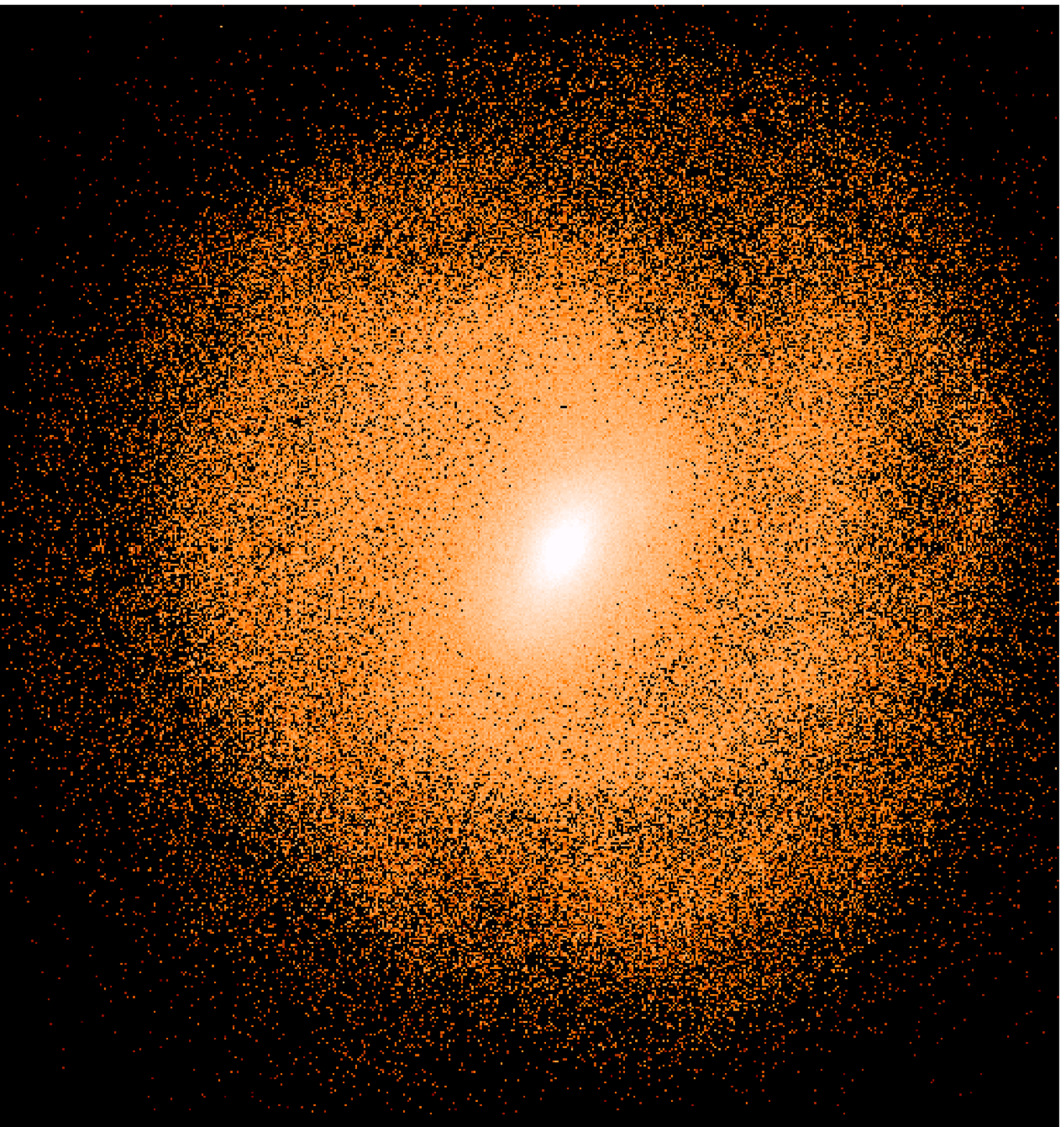}
\includegraphics[scale=0.22]{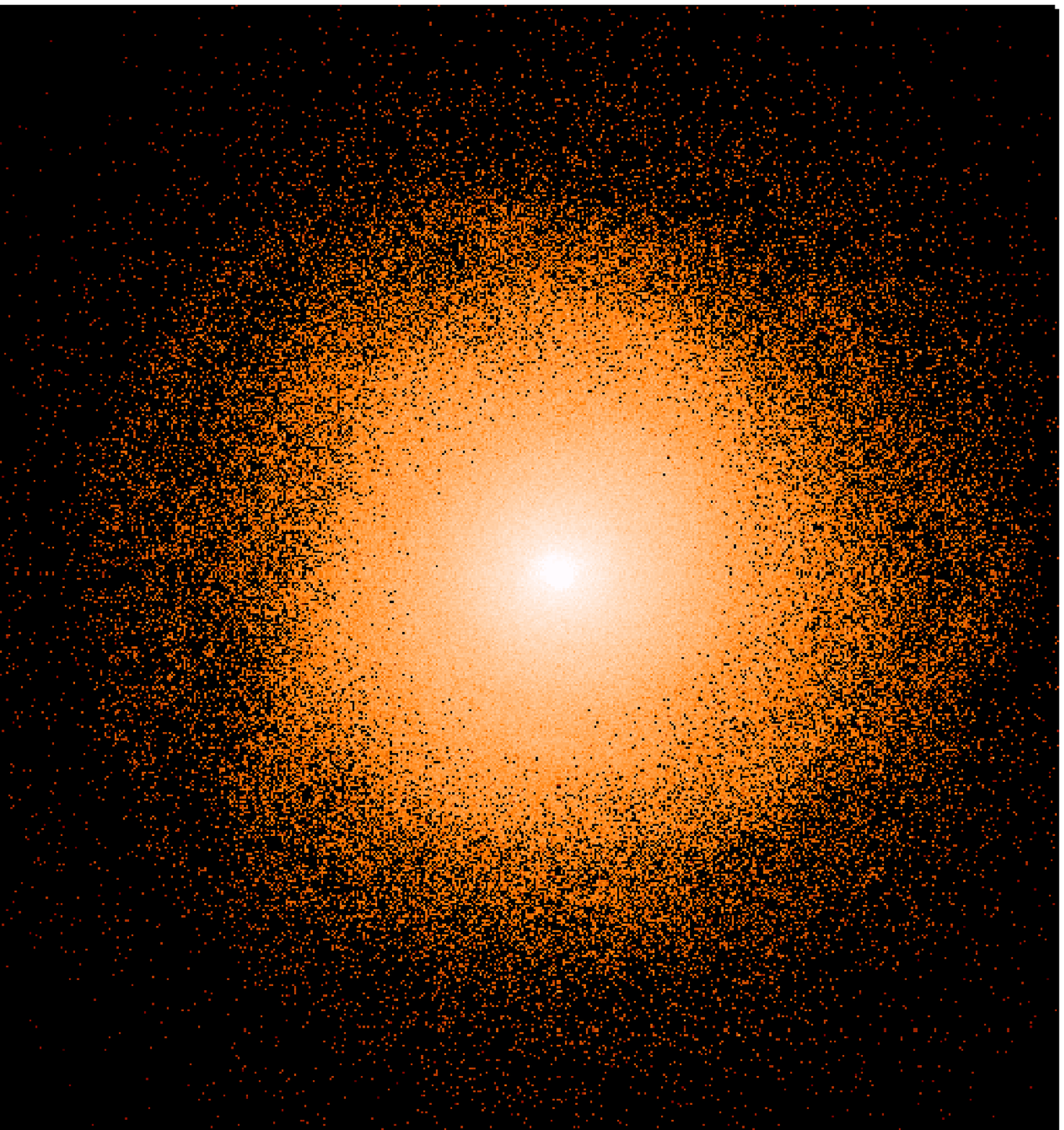}
\caption{Distribution of stars for galaxy H barred (left column) and 
unbarred (right column), at $t=0.4$ (top), 0.8 (middle), and $1.2\,\rm Gyr$
(bottom). Each panel covers $30\times30\,\rm kpc$.}
\label{fig:H}
\end{figure}

\subsubsection{Bar Strength}

Figure~\ref{fig:H} shows the distribution of
stars at three different times, for galaxy H (barred and unbarred).
At $t=0.4\,\rm Gyr$ (top left panel),
the bar and spiral arms are clearly visible in the barred galaxy.
Visually, the length and ellipticity of the bar 
appears to remain roughly constant up to
$t=1.2\,\rm Gyr$ (bottom left panel), while the spiral pattern is getting
more diffuse. By contrast, the unbarred galaxy shows hardly any structure.

Bar strength has been shown to be a particularly important parameter in the
evolution of barred galaxies, having a major impact on both the SFR history
and the gas mixing \citep{athanassoula_angular_2003,buta_distribution_2005,
hoyle_galaxy_2011,scannapieco_observers_2010,wang_quantifying_2012}. Several
different definitions have been proposed to quantify bar strength, either
through angular momentum transfer \citep{buta_distribution_2005,grandetal12},
ellipse-fitting, or Fourier analysis \citep{aguerri_population_2009}. In
this paper we calculate bar strength using a method proposed by
\citet{athanassoula_morphology_2002}, based on the components of the Fourier
decomposition of the azimuthal distribution of particles. The
components $a_m$ and $b_m$  are given by :
\begin{eqnarray}
a_m(R)&=&\sum^{N_R}_{n=1} \cos(m \theta_n),\, m=0,1,2,\ldots;\\
b_m(R)&=&\sum^{N_R}_{n=1} \sin(m \theta_n),\, m=0,1,2,\ldots;
\end{eqnarray}

\noindent
where $N_R$  is the number of particles within a radius $R$ and $\theta_n$
is the azimuthal angle of the particle $n$. We compute $a_2(R)$ and $b_2(R)$
for different radii $R$ and define the bar strength $A_2$ as 
\begin{equation}
 A_2=\max A_2(R)=\max \left( \frac{\sqrt{a_2^2+b_2^2}}{a_0}\right).
\end{equation}

\begin{figure}
\centering
\includegraphics[scale=0.41]{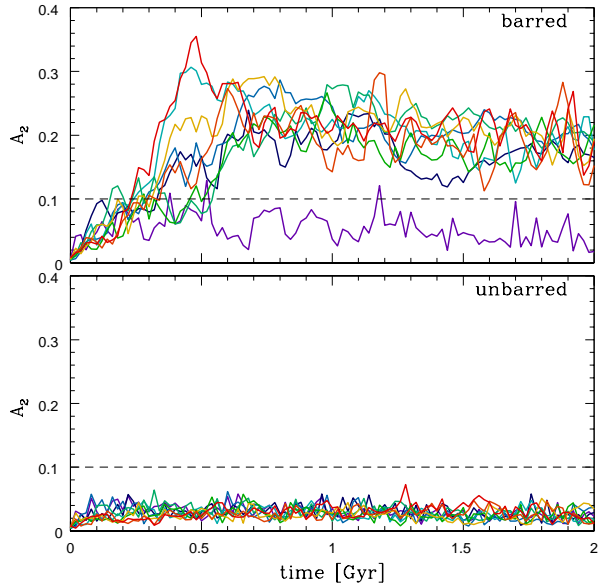}
\caption{Bar strength as a function of time for all galaxies. Top panel: 
barred galaxies. Bottom panel: unbarred galaxies. 
Dashed lines indicate the value $A_2=0.1$ used to identify bars. 
Colour coding follows 
the one in Table \ref{table:initial}. }
\label{fig:str}
\end{figure}

This is the amplitude of the $m=2$ mode, normalized to the mean density.
In Figure \ref{fig:str} we present the evolution of the bar 
strength as a function of time for both barred and unbarred galaxies. 
The top panel shows that most galaxies we 
consider as barred have an average 
bar strength of $A_2=0.125$ at all times past $t=0.5\,\rm Gyr$. 
Galaxies H and D 
have a faster-than-average increase and higher magnitude of their bar 
strength, peaking at almost $A_2=0.2$ at $t=0.45\,\rm Gyr$, but their strength 
fall off to the average value within $200\,\rm Myr$. 
The exception is the lower-mass galaxy O, which has no significant bar. 
Generally speaking, all our barred 
galaxies with the exception of galaxy~O have similar 
bar strength values and history.

In contrast to barred galaxies, all our galaxies marked as unbarred do not
have any significant azimuthal periodicity, indicating that none of them
have anything close to a clear bar structure. 
In the reminder of this paper, we will use
$A_2=0.1$ as the transition between barred and unbarred galaxies. Hence,
in a situation when $A_2$ increases with time,
the value $A_2=0.1$ corresponds the
onset of bar formation. This choice is somewhat arbitrary, but not critical
as the value of $A_2$ tends to increase rapidly during bar formation.

\begin{figure}
\centering
\includegraphics[scale=0.41]{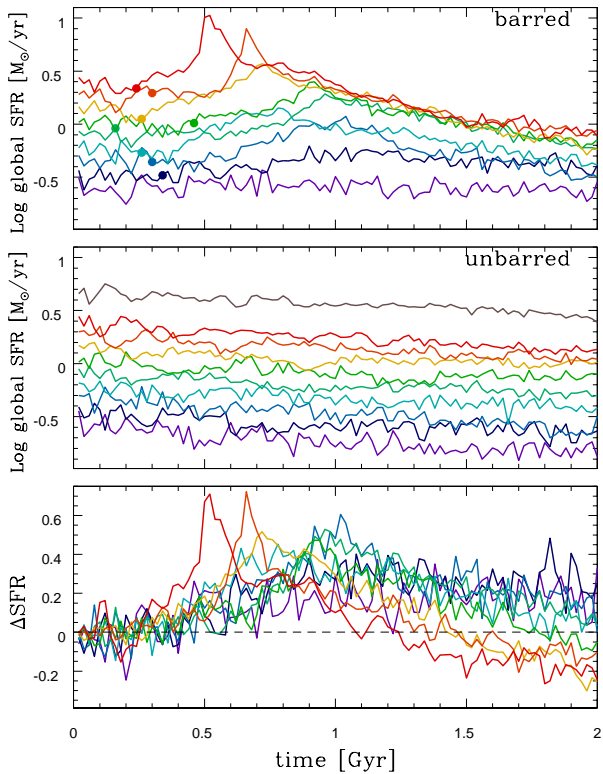}
\caption{Time-evolution of the global star formation rate for all simulated 
galaxies. Top panel: barred galaxies; middle panel: unbarred galaxies; bottom 
panel: difference between barred and unbarred galaxies. Big dots in top panel 
show the time when the bar appears, defined by  $A_2=0.1$. The O galaxy's bar 
is never strong enough to classify its host as barred under this criteria 
(bottom, purple curve it top panel).}
\label{fig:SFRG}
\end{figure}

\subsubsection{Star Formation Rate}

Figure~\ref{fig:SFRG} shows the global SFR in barred and unbarred galaxies as
a function of time as well as the difference 
$\Delta{\rm SFR}=\log({\rm SFR_{bar}})-\log({\rm SFR_{nobar}})$
between them. 
The SFR in barred galaxies (top panel) remains constant
or decreases slowly until the bar forms
somewhere between $t=0.3$ and $0.6\,\rm Gyr$ (big dots in top panel show the
point where the bar strength $A_2=0.1$, defined as the onset of bar 
formation). Once the bar is present, high-mass galaxies 
experience a rapid increase in star formation, which we shall 
refer to as a {\it starburst\/},
with the SFR reaching its maximum value somewhere between 0.5 and $1\,\rm Gyr$.
Among these high-mass galaxies, the highest-mass ones (F-G-H) 
not only show a much stronger increase
of their SFR, reaching over 400\% of the initial levels, but the SFR peaks
earlier, between 0.5 and $0.7\,\rm Gyr$. In contrast, 
the low-mass galaxies A-B-C show a gentler and slower increase in 
their SFR, peaking around 
$t\approx1\,\rm Gyr$ with a SFR about twice as large as the initial one.
After the peak is reached, the SFR decreases with time at a rate
that is roughly the same for all galaxies.
Note that in the lowest-mass galaxy O, the bar never formed, and the
SFR slowly decreases with time.

The SFR in unbarred galaxies (middle panel)
varies very smoothly. It either remains constant
or decreases slowly, dropping by a factor of 2 over $2\,\rm Gyr$. 
As expected, both sets of galaxies show very similar SFRs until the bar forms.
The lower panel of Figure~\ref{fig:SFRG} shows the
difference $\Delta$SFR between barred and unbarred galaxies.
Star formation tends to be significantly enhanced by the presence
of a bar. Also, notice that even though barred galaxy O does not reach a 
bar-qualifying 
$A_2$ value, it still has a higher SFR than its unbarred counterpart, 
implying that even 
modest deviations from axisymmetry can be sufficient to feed gas inflows
toward the centre. While $\Delta$SFR remains positive at low and
intermediate masses, it eventually becomes negative at high-masses
(galaxies E-F-G-H). The rapid decreases in SFR after
the starburst results in $\Delta$SFR becoming negative at late times.
The post-starburst decrease in SFR in massive 
barred galaxies is caused by gas depletion, as we showed in Paper~I. 
Recall that all of our galaxies are isolated simulations that are not 
embedded in a cosmological context.
Accretion and mergers could potentially replenish the supply of gas,
affecting the late-time evolution of the galaxy.
We will investigate these processes in future work.
For now, we will simply remember that the post-starburst SFR 
might not apply to all barred galaxies, and should be regarded as a 
lower limit. 

\subsection{Central properties}

We now focus on the dynamics of the central region of our galaxies,
which we define as a 1 kpc radius cylinder, parallel to the rotation axis of
the galaxy and aligned with the center of mass of the stellar component.
This central region is where most of the bar-driven gas accumulates, and
roughly corresponds to the zone covered by an optical fibre of the SDSS. For
redshifts $0.02<z<0.1$, a 3 arcsecond fibre diameter corresponds to a
physical radius of $0.61-2.76\,\rm kpc$ 
(for a concordance $\Lambda$CDM model with $\Omega_0=0.275$,
$\lambda_0=0.725$, $h=0.702$). 
Gas and stars located inside the bar move along 
elongated orbits, entering and exiting the central region, and these orbits
evolve as angular momentum is being redistributed.
The collisionless stellar component of the bar 
stabilises after bar formation (see Fig.~\ref{fig:H}),
while the orbits of the gas elements contract 
with time. This causes a net increase of the gas mass inside the central 
region, as more gas moves in than moves out. Eventually,
the gaseous component of the bar is entirely contained inside 
the central region, and from that moment
this region behaves as an isolated system, with negligible gas flows
across its boundary. 
In Table~\ref{table:final}, we give 
the stellar mass $M_*$ in the central region 
at the end of the simulations. Not surprisingly, the numbers increase with
increasing initial stellar mass, and are larger for a barred galaxy than for
an unbarred one with the same initial stellar mass. The unbarred galaxy H
has a final central stellar mass of $4.25\times10^9\msun$, similar to 
barred galaxy E. This was our rationale for simulating the high-mass
unbarred galaxy I, to provide a basis for comparison with the massive
barred galaxies F, G, and H.


\begin{table}
 \centering
 \begin{minipage}{80mm}
  \caption{Stellar mass $M_*$ in the central $1\,\rm kpc$
region at $t=2\,\rm Gyr$, in units of $10^9\msun$}
  \begin{tabular}{@{}ccc@{}}
  \hline
Galaxy & Barred & Unbarred \\
  \hline
O & 1.14 & 0.84 \\
A & 1.73 & 1.04 \\
B & 2.21 & 1.28 \\
C & 2.77 & 1.58 \\
D & 3.58 & 1.96 \\
E & 4.12 & 2.35 \\
F & 5.55 & 2.89 \\
G & 6.63 & 3.55 \\
H & 8.89 & 4.25 \\
I & $\cdots$ & 7.67 \\
   \hline
\end{tabular}
\label{table:final}
\end{minipage}
\end{table}

\begin{figure}
\centering
\includegraphics[scale=0.41]{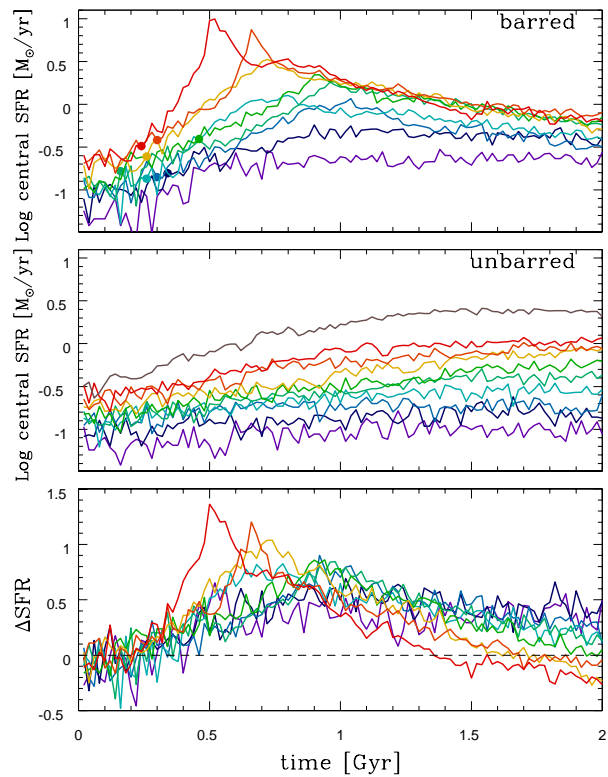}
\caption{Time-evolution of the central star formation rate for all simulated
galaxies. Top panel: barred galaxies; middle panel: unbarred galaxies;
bottom panel: difference between barred and unbarred galaxies. Big dots
in top panel show the time at when the bar appears, defined by $A_2=0.1$}
\label{fig:SFRF}
\end{figure}

We plot the SFR within the central region as a function of time in 
Figure~\ref{fig:SFRF} for barred and unbarred galaxies, as well as the 
difference $\Delta$SFR between them. All barred
galaxies start with a relatively low SFR until the bar 
forms. Then, the net influx of gas caused by 
the bar 
increases the gas mass in the central region, with a corresponding increase 
in SFR. As for the global SFR, the central SFR peaks between $0.5$ and 
$\rm 1\,Gyr$, depending on the mass of the galaxy, with high-mass galaxies 
peaking earlier than low-mass ones. The SFR then decreases smoothly for the 
massive galaxies while remaining almost constant in the low-mass ones such 
as O and A. When compared with global SFR (Fig.~\ref{fig:SFRG}), central SFR 
starts out much lower but it rises very fast in barred galaxies, dominating 
the global behaviour by $t=0.7\rm\,Gyr$.
In contrast, the evolution of central SFR of unbarred galaxies is much
smoother. While there is a slight increase with time,
the amplitude and fluctuations of the SFR are much smaller than the ones
in barred galaxies. 

\begin{figure}
\centering
\includegraphics[scale=0.41]{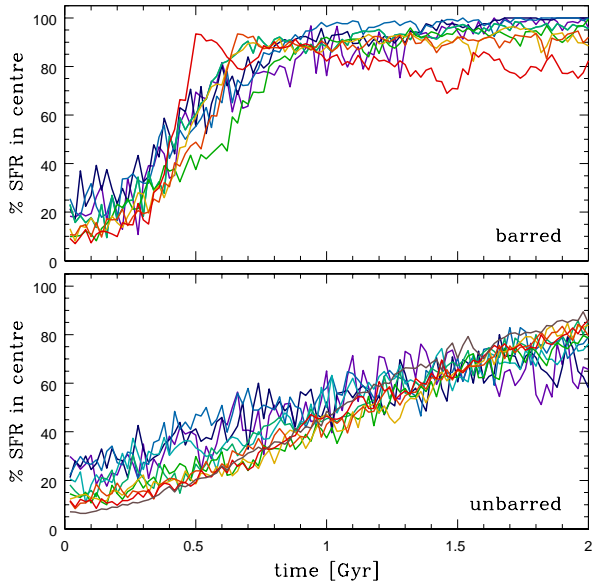}
\caption{Percentage of the star formation of the galaxy occurring within the 
$1\rm\,kpc$ central region. Top panel: barred galaxies; bottom panel: 
unbarred galaxies.}
\label{fig:SFRpercent}
\end{figure}

To contrast the evolution of the global and central SFR,
we plot in Figure~\ref{fig:SFRpercent} the fraction of star 
formation happening in the central $1\,\rm kpc$ region.
All barred galaxies follow a similar evolution, independently 
of their stellar mass. During the first $0.2\,\rm Gyr$,
low-mass galaxies have a higher fraction of their global SFR going 
on in the central region because this region 
covers a greater fraction of the total area 
of low-mass galaxies than it does in high-mass galaxies. 
At $t=0.3\rm\,Gyr$, the bar has formed, and star formation
becomes more dominated by activity in the central region. In less than
$0.5\rm\,Gyr$ more than 75\% of the galaxy's new stars are formed in the 
central region.
All unbarred galaxies also follow a similar evolution, independently 
of their stellar mass. During the first $0.2\,\rm Gyr$, the SFR
follows the one in barred galaxies
because the bar in those galaxies has not formed yet.
Afterward, the SFR increases smoothly, and
star formation is concentrated predominantly in the central region 
at late times. While the fraction steadily
increases to 75\% through the simulation, it does so in a gentle manner,
suggesting a smooth and continuous inflow of gas instead of massive one
as the ones in barred galaxies.

\begin{figure}
\centering
\includegraphics[scale=0.41]{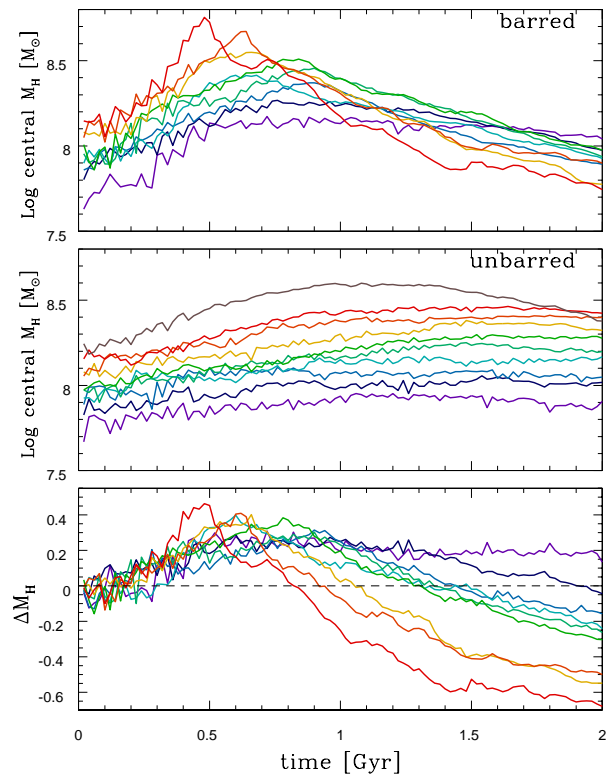}
\caption{Time-evolution of gaseous hydrogen mass in the central region for each
galaxy. Top panel: barred galaxies; middle panel: unbarred galaxies;
bottom panel: difference between barred and unbarred galaxies.}
\label{fig:MHF2}
\end{figure}

Figure \ref{fig:MHF2} shows the evolution of the hydrogen mass
$M_{\rm H}$ inside the
central region as a function of time.
In barred galaxies, gas flows along
the bar during the first $0.5-0.7\,\rm Gyr$, increasing the central value 
of $M_{\rm H}$. Then, star formation consumes the gas,
reducing the value of $M_{\rm H}$.
The gas consumption varies greatly with the total stellar
mass: the central region 
of high-mass galaxies first gets a higher and faster influx
of gas as the bar forms, but it also consumes a much greater fraction of the
gas with time, depleting the central region 
of gaseous content. In the extreme cases
of galaxies A (black) and H (burgundy), galaxy A ends up with more 
hydrogen gas in the central region 
than galaxy H, even though galaxy H starts with three times as 
much gas. 

The results shown in Figure~6 elucidate the differences in SFR as a function of 
$M_*$ seen in Figures~3 and~4. High-mass
galaxies get a huge amount of gas driven in their core by the bar which
cause a starburst-like increase in the global
SFR. After $0.5-0.7\,\rm Gyr$, the
bar has already moved most of the gas inside the galactic core, leaving
the galaxy partially gas-depleted, which then
brings the SFR down. Low-mass
galaxies do not move enough gas to create a starburst; most of the gas also
flows in the central region 
within the first $1\,\rm Gyr$ but it accumulates in the
central region and is slowly transformed into stars, leaving a sightly
decreasing amount of gas and a corresponding flat SFR.
To illustrate the accumulation of gas in the center of unbarred galaxies, we
plot in the bottom panel of Figure~\ref{fig:MHF2} the difference
$\Delta M_{\rm H}=\log(M_{\rm H,bar})-\log(M_{\rm H,nobar})$. The values initially 
increase, as barred galaxies drives gas efficiently toward the center.
Then, the starburst taking place in barred galaxies
results in a large consumption of gas, and eventually $\Delta M_{\rm H}$
becomes negative at all masses except the lowest one. The effect 
is particular strong at high masses (galaxies F, G, and G), where the
starburst is the strongest. 

\begin{figure*}
\centering
\includegraphics[scale=0.8]{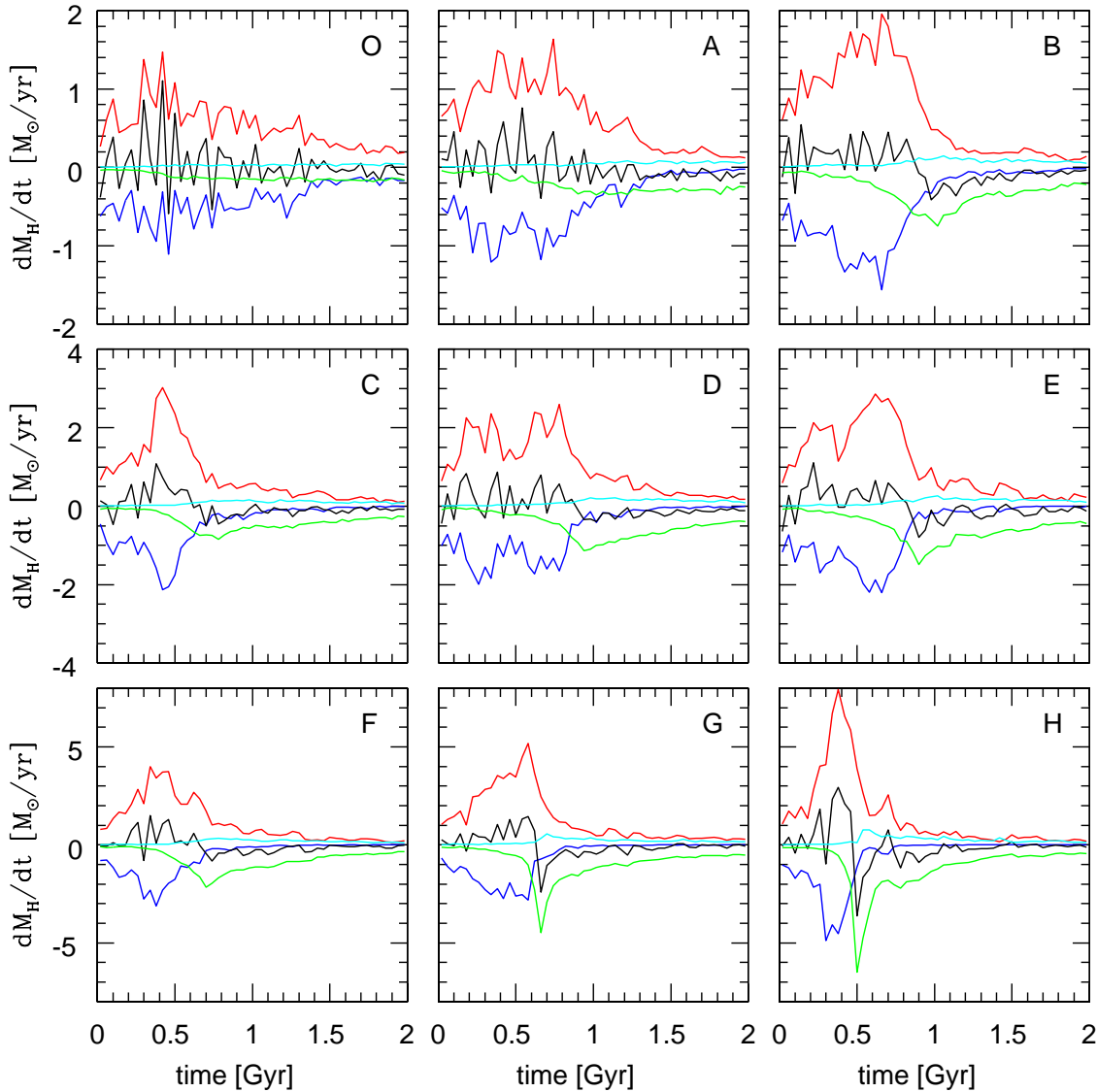}
\caption{Rate of change of the hydrogen mass within the central region 
for each
barred galaxy. Gas motion is in red for inflows and blue for outflow;
stellar formation gas consumption is in green and the supernovae
feedback is in cyan. Black line represents the net changes.}
\label{fig:inflowbar}
\end{figure*}

We verify our predictions about gas dynamics in the 
central region by analyzing the
physical effects responsible for changing the gaseous mass within the
central region,
as we did in Paper I. Three physical processes are crucial here: First, as
gas moves along elliptical orbits along the bar, it flows in and
out of the central region,
increasing and decreasing the central gas mass. 
Secondly, star formation decreases the gas mass as gas 
is converted into stars. Finally, stellar evolution feedback returns gas
to the ISM through supernovae and stellar winds. Figure \ref{fig:inflowbar}
shows these three processes in all barred galaxies: gas moving into the
central region
is shown in red, gas moving out in blue, star formation in green,
stellar feedback in cyan, and the total effect in black. Gas flow dominates
early on as the gas moves along elongated orbits that
cross the boundary of the central region.
However, as the bar transfers angular momentum away from the gas,
the orbits become smaller, and once they become smaller than the
central region there is no longer any gas flow across the boundary
of the central region. We see in all
galaxies except O that there is a time between $t=0.6\,\rm Gyr$ and 
$t=1.2\,\rm Gyr$ where almost
all the gas is trapped within the central region, 
and that this gas
is then consumed through an important star formation boost. The relative
importance of the SFR-related decrease in hydrogen mass (green lines
in Fig.~\ref{fig:inflowbar}) 
is greater in high-mass galaxies than in low-mass ones.

This high gas consumption in high-mass barred galaxies is
highlighted by the inversion of the mass lines after $1.5\,\rm Gyr$ 
in the top panel of
Figure~\ref{fig:MHF2}: The lowest-mass galaxies are the ones with the highest
central hydrogen mass, while the highest-mass galaxies have the lowest one.
Comparing the bottom panels of Figures~\ref{fig:SFRF} and~\ref{fig:MHF2},
we see that at late time ($t=2\,\rm Gyr$), 
barred galaxies maintain a higher SFR than 
unbarred galaxies while having a smaller gas supply. This indicates that
the enhancement in central SFR in barred galaxies is not caused by a larger gas
supply, but instead by a higher star formation efficiency. This
is in agreement with the recent results of \citet{sandstrometal16}.

The dramatic late-time reduction in $M_{\rm H}$ for 
massive barred galaxies is caused
by gas exhaustion. Massive amounts of gas are consumed during the starburst
in these galaxies, limiting their ability to accumulate gas in the central
region as unbarred and less-massive barred galaxies do. Because our
simulations do not include gas accretion from the IGM, or galaxy mergers,
the gas supply is limited. Accretion could enable massive barred galaxies
to sustain a high SFR for a longer period, or it could make the starburst 
even stronger. 

By comparing barred and unbarred galaxies with the same global stellar mass,
\citet{ellison_impact_2011} found an enhancement in central SFR for barred
galaxies with global $M_*>10^{10}\msun$, with no corresponding enhancement
in central metallicity. These authors suggested a possible explanation
based on the simulations of \citet{combes_bars_1993}. These simulations showed
that in low-mass, late-type galaxies, bars stop growing at an early stage,
while bars in high-mass, early-type galaxies grow continuously. Since bars
are responsible for driving gas toward the central region and fueling
central star formation, star formation would be
ongoing in high-mass barred galaxies 
and not in low-mass ones. 
However, our simulations show that star formation is sustained in both 
unbarred and barred galaxies. Indeed, in unbarred galaxies and in low-mass
barred ones, it is the low
star formation efficiency (compared to massive barred galaxies)
that enables these galaxies to
sustain star formation for a long period of time
without exhausting the gas supply.

There is an important effect that we must be aware of when
comparing barred and unbarred galaxies with a same central
$M_*$. Two galaxies with different total stellar masses can have the
same central $M_*$ if they happen to be at different stages of
their respective evolution. We will now address this particular issue.

\subsection{Barred vs unbarred galaxies as a function of mass}

\subsubsection{Galaxy Samples}

While studying the evolution of barred and unbarred galaxies at different 
times gives us a very good insight on the different dynamics ruling both of 
them at different mass scales, this kind of direct temporal comparison is far 
from ideal when it comes to understanding observational results. Instead, we 
need to recreate such results with variables that are observationally 
available and then interpret them with our temporal and evolutionary 
knowledge, as done in \citet{scudder_galaxy_2015}. 

\begin{figure}
\centering
\includegraphics[scale=0.41]{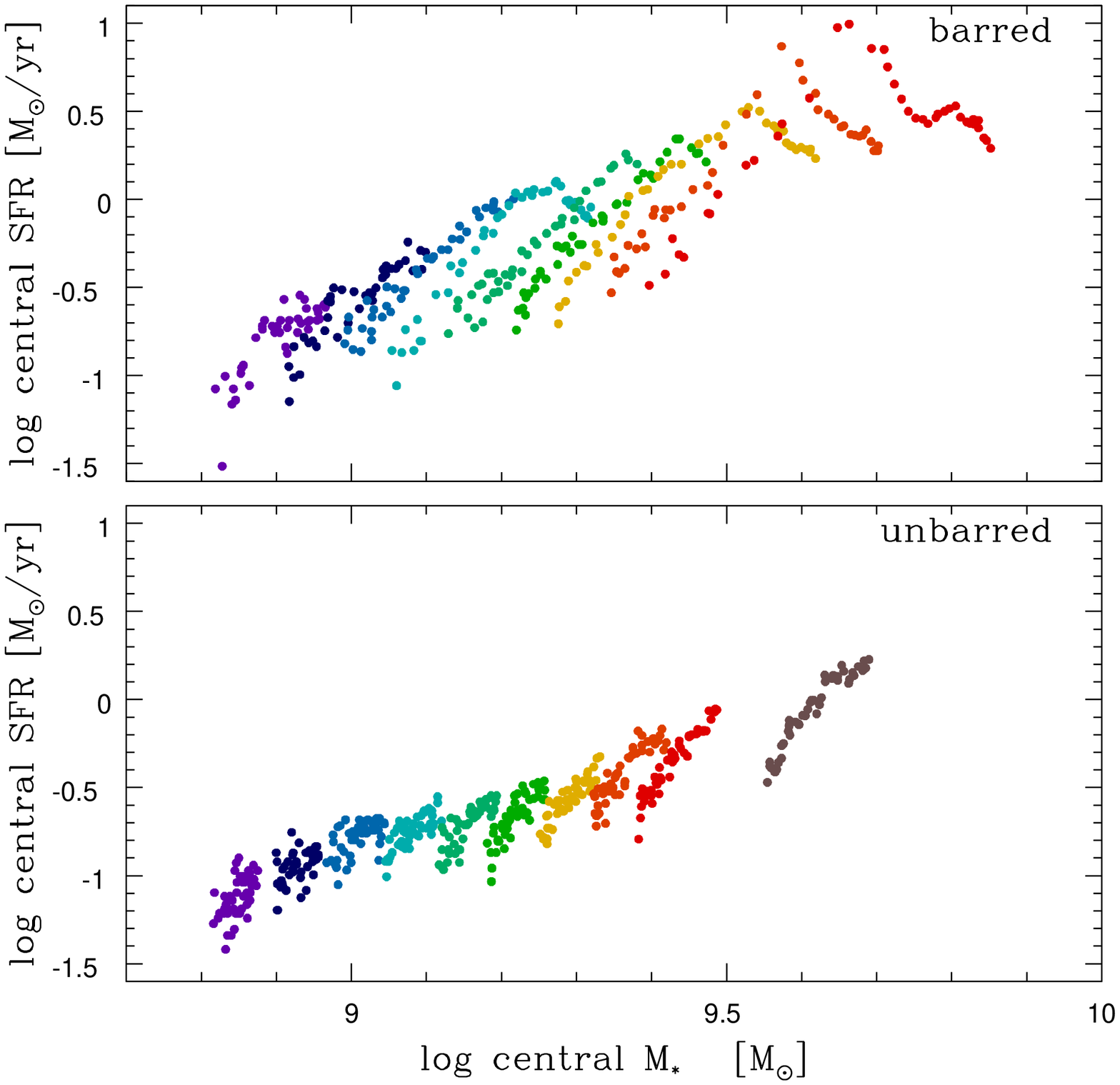}
\caption{SFR as a function of the stellar mass, 
in the central region. Each dot corresponds to a snapshot of a simulation.
Colours indicate the corresponding galaxies.}
\label{fig:SFRFm}
\end{figure}

\begin{figure}
\centering
\includegraphics[scale=0.41]{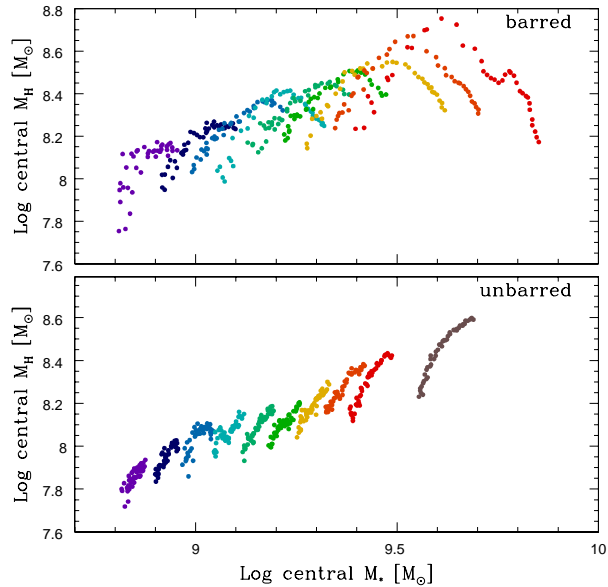}
\caption{Hydrogen mass as a function of the 
stellar mass, in the central region. 
Each dot corresponds to a snapshot of a simulation.
Colours indicate the corresponding galaxies.}
\label{fig:MHFm}
\end{figure}

In Figures~\ref{fig:SFRF} and \ref{fig:MHF2}, 
we showed the time-evolution of the SFR and hydrogen
mass in the central region. 
With real, observed galaxies, we do not have a 
direct measure of the time $t$, so we must rely on another observable
to identify the various evolutionary stages of the galaxies. A good
choice is the central stellar mass $M_*$, which increases monotonically
with time.
We plot the SFR and hydrogen mass $M_{\rm H}$ in the central region 
as a function of the central 
stellar mass in Figures \ref{fig:SFRFm} and \ref{fig:MHFm}, respectively.
Each dot represents one snapshot of a simulation. At the beginning 
of each simulation, it takes a certain amount
of time before the gas distribution
relaxes to equilibrium, so we excluded the first $100\,\rm Myr$ 
from the analysis. 
For barred galaxies, we also exclude all the snapshots that precede
the bar formation, as defined by $A_2=0.1$. 
At $t>{\rm1\,Gyr}$, the barred galaxies are in their post-starburst
phase, where accretion could have an impact. Hence, for this analysis,
we are only including galaxies at $t<{\rm1\,Gyr}$, which roughly corresponds 
to their period of high activity shown in 
Figure \ref{fig:inflowbar}. 

In the bottom panel of Figure \ref{fig:SFRFm}, we see that unbarred galaxies of
different masses combine to form a tight relation between SFR and
central stellar mass, especially if we neglect the early stages of
evolution (leftmost part of each curve). This relation is well-approximated
by a power law, ${\rm SFR}\propto M_*^\alpha$, with $\alpha\sim1.35$.
In the upper panel, we see that barred galaxies behave very differently.
The SFR of a galaxy of a given total mass does not vary monotonically, and
covers a wide range of values compared to unbarred galaxies of the same mass.
Also, when combining barred galaxies of different masses, they do not form
a single relation, unlike unbarred galaxies. Overall, the SFR tends to
increases with central stellar mass, and roughly follows the same power 
law as for unbarred galaxies, but there is a lot of scatter.
Clearly, there is a risk of comparing galaxies of completely different 
global $M_*$ that happen to
have the same central mass because they are at different 
stages of their respective evolution. For example, 
at $\log M_*=9.4$ we find the H galaxy in its pre-bar stage,
the C galaxy in its final, post-burst stage, the F galaxy in the middle 
of its gas inflow period, and the E at the peak of the SFR. 
  
The relation between central $M_{\rm H}$ and central $M_*$ is shown in 
Figure \ref{fig:MHFm}. The results are similar to the ones shown in
Figure \ref{fig:SFRFm}. Central $M_{\rm H}$ 
increases monotonically with central $M_*$ in unbarred galaxies.
Combining all unbarred galaxies of different masses, we find again
a power law, $M_{\rm H}\propto M_*^\alpha$, with $\alpha\sim0.75$.
In contrast, central $M_{\rm H}$ does not vary monotonically with central $M_*$
for barred galaxies of a given total mass, and when combining all barred 
galaxies, there is very little correlation between central $M_{\rm H}$ and 
central $M_*$.
Since, in each barred galaxy, 
central $M_{\rm H}$ first increases with time because of gas inflow, 
then decreases because of star formation while central $M_*$ steadily 
increases, there are usually two very different values of $M_*$ 
corresponding to a given $M_H$. What correlates well with $M_*$
is the peak value of the SFR in the central region, 
but determining observationally that the SFR 
is at its peak value is not possible.

\subsubsection{Mass binning}

\begin{figure}
 \centering
 \includegraphics[scale=0.41]{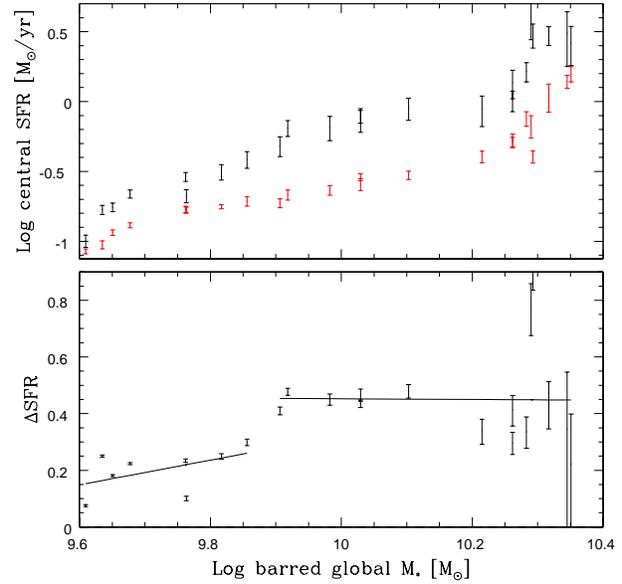}
  \caption{Average SFR (top panel) and $\Delta$SFR (bottom panel) in each
central mass bin as a function of the barred galaxy $M_*$. Black and red symbols
in upper panel represent barred and unbarred galaxies respectively. 
Lines are the best linear 
fitting on galaxies of masses lesser and greater than $\log M_*=9.9$}
  \label{fig:same_fiber_young}
\end{figure}

To study the general dependence of SFR with central $M_*$, we must acknowledge 
that galaxies of different global $M_*$ can have the same central $M_*$ because
they are at different evolutionary stages. As such, to compare galaxies of 
a given central $M_*$, we must account for both high-global-mass, young 
galaxies and lower-global-mass but older galaxies having the same central
$M_*$. \citet{ellison_impact_2011} obtained their two-regime relation by 
binning 
their sample of 294 barred galaxies depending on the stellar mass in the
central region. Once binned, they compared the averaged SFR in each bin to 
the expected SFR of unbarred galaxies of corresponding central mass and 
noted that only high-mass galaxies have a larger central SFR. 
To recreate 
a similar method, we consider the various dots in 
Figures~\ref{fig:SFRFm} and \ref{fig:MHFm} as
representing different galaxies at the present, instead of a few galaxy at 
several different times.
This gives us a sample of several hundreds barred and unbarred galaxies.
We then calculate a weighted average SFR and weighted average global 
stellar mass in 30 central mass bins for barred and unbarred galaxies.
Weights are selected to represent the relative likelihood of observing a 
particular galaxy. To determine the weights, 
we use the halo mass function of \citet{murray_hmfcalc:_2013}.
Then, following \citet{ellison_impact_2011}, we calculate 
$\Delta{\rm SFR}$ for each 
mass bin and plot it as a function of the
global $M_*$ of the barred galaxy. 
The results are shown in Figure~\ref{fig:same_fiber_young}.
Low-mass barred galaxies have a small
enhancement of 0.2 dex of their central SFR when compared to unbarred 
galaxies with equivalent central mass. This enhancement quickly increases 
and stabilize around 0.4 dex, with the transition being centered at 
$\log M=9.9$. 

\begin{figure}
\centering
\includegraphics[scale=0.41]{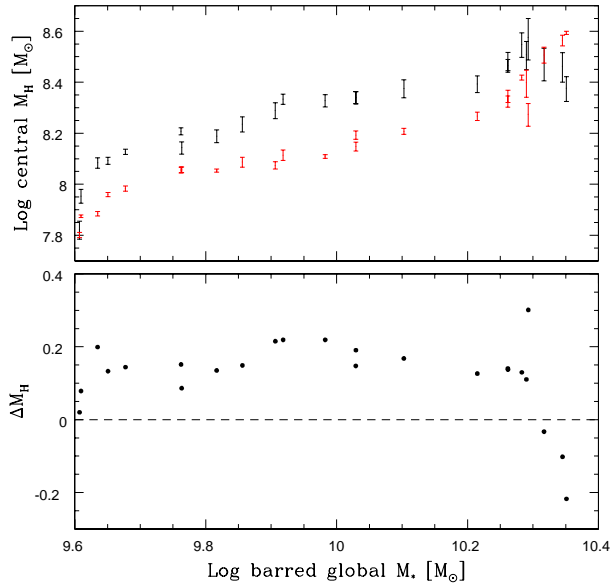} 
\caption{Average $M_{\rm H}$ (top panel) and $\Delta M_{\rm H}$ 
(bottom panel) in each
central mass bin as a function of the barred galaxy $M_*$. Black and red symbols
in upper panel represent barred and unbarred galaxies respectively.}
\label{fig:delta_H}
\end{figure}

In the top panel of Figure~\ref{fig:delta_H}, we plot the central hydrogen mass 
for barred and unbarred galaxies. The values are fairly constant, even for
barred galaxies. Higher-mass barred galaxies drive more gas toward the
central region, more of that gas is converted into stars. However, at
$\log M_*=10.3$, there is both a sudden drop in $M_{\rm H}$ for barred
galaxies and a sudden increase for unbarred galaxies. The bottom panel
of Figure~\ref{fig:delta_H} shows the difference
$\Delta M_{\rm H}$. At $\log M_*<10.3$,
$M_{\rm H}$ is larger by $0.2\,\rm dex$ for barred galaxies. At
$\log M_*=10.3$, $M_{\rm H}$ suddenly drops by $0.4\,\rm dex$ down to
negative values. This bin is dominated by high-mass barred galaxies
in their starburst phase, where gas consumption is very efficient
compared to unbarred galaxies.

\subsection{Initial gas fraction}

\begin{figure}
 \centering
 \includegraphics[scale=0.41]{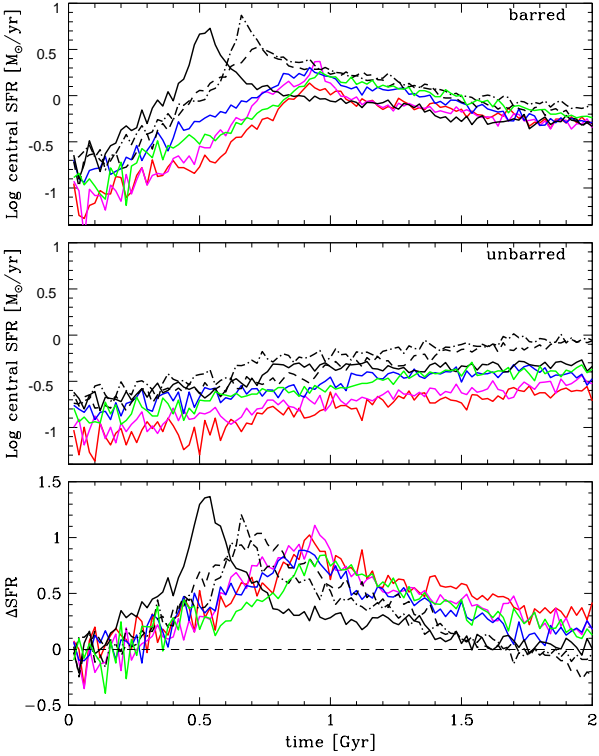} 
  \caption{Time-evolution of the central SFR for the D galaxies set.
Top panel: barred galaxies; bottom panel: unbarred galaxies. Colour 
coding for the solid lines follows the one in Table~\ref{table:gas}.
Dashed and dot-dashed black lines show respectively galaxies F and G, 
which are taken from the original set of simulations
(see Table~\ref{table:initial}).}
 \label{fig:SFRF_g}
\end{figure}

\begin{figure}
 \centering
 \includegraphics[scale=0.41]{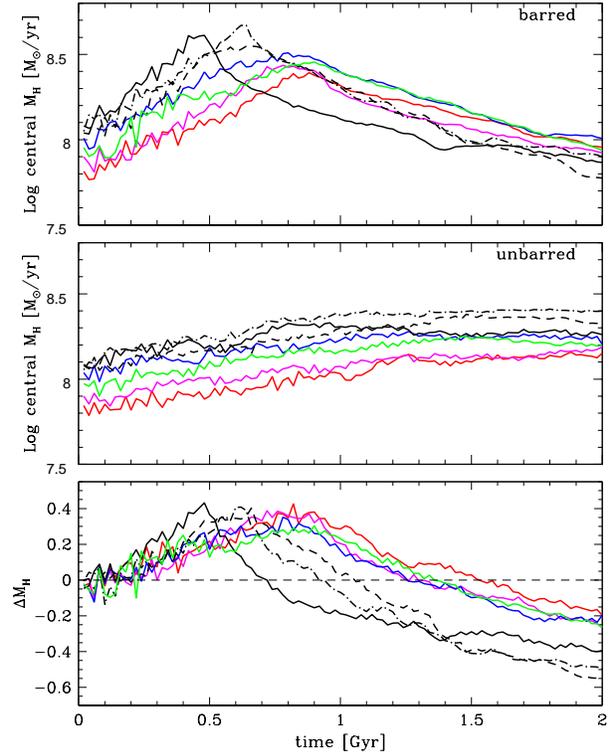} 
  \caption{Time-evolution of gaseous hydrogen mass in the central region 
for the D galaxies set. 
Top panel: barred galaxy; bottom panel: unbarred galaxies. Dashed
and dot-dashed lines show galaxies F and G, respectively.}
  \label{fig:MHF_g}
\end{figure}

While we have been comparing the evolution of our galaxy set focusing on the
stellar mass of the galaxy, this is not the only variable parameter in our
set: the total amount of gas available is another important one which
can greatly impact the star formation rate. 
To explore how this affects our results, we
performed variations of the run D: two with a lower initial gas fraction
($\rm D^-$ and $\rm D^{--}$) and two with a higher initial gas fraction 
($\rm D^+$ and $\rm D^{++}$); all of these simulations where performed both 
in barred and unbarred galaxies for a total of 8 new simulations. Details 
of their gaseous component are given in Table \ref{table:gas}. The last 
column refers to the colours used in Figures \ref{fig:SFRF_g} and 
\ref{fig:MHF_g}.


\begin{table}
 \centering
 \begin{minipage}{80mm}
  \caption{Initial proprieties of the D galaxies. Masses are in
units of $10^9\msun$}
  \begin{tabular}{@{}lccccl@{}}
  \hline
Galaxy & $M_*$ & $M_{\rm gas}$ & $N_{\rm gas}$ & $f_{\rm gas}$ & colour\\
  \hline
$\rm D^{--}$ &10.0 &2.50 &34 717 & 0.200  & red \\
$\rm D^-$   &10.0 &2.98 &41 480 & 0.229 & magenta \\
D           &10.0 &3.51 &48 749 & 0.259 & green \\
$\rm D^+$   &10.0 &4.10 &56 943 & 0.290 & blue\\
$\rm D^{++}$ &10.0 &4.70 &65 350 & 0.319 & black\\
   \hline
\end{tabular}
\label{table:gas}
\end{minipage}
\end{table}

Figure \ref{fig:SFRF_g} shows the SFR inside the
central region of the various D 
galaxies as a function of time. Barred galaxies from $\rm D^{--}$ to $\rm D^+$
all have a very similar SFR history: their SFR increases steadily from 0 to 1 
Gyr before decreasing, in a very similar way to other intermediate-mass 
galaxies (C to E). However, the barred $\rm D^{++}$ galaxy shows a large 
SFR peak around $t=0.5\,\rm Gyr$ 
much more akin to the high-mass galaxies F-G-H. 
In Figure \ref{fig:MHF_g}, we show that the amount of gas present in the 
barred central region behaves quite differently for the $\rm D^{++}$
galaxy compared to the less gaseous ones: in the first four, the amount of
gas varies smoothly and the galaxies with the most gas always have the most
gas. However, in $\rm D^{++}$ there is an important increase in the central
gas mass during the first 0.5 Gyr, before falling down due to the
starburst-like formation episode. Unbarred galaxies do not show such
changes in high-gas-mass galaxies.

Barred galaxies $\rm D^{++}$ and F have similar initial gas mass 
($M_{\rm gas}=4.0\times10^9\msun$, vs. $M_{\rm gas}=5.02\times10^9\msun$,
a difference of 6\%). On the top panels of Figures~\ref{fig:SFRF_g}
and~\ref{fig:MHF_g}, we added the results for galaxy F (dashed lines), 
and also galaxy G (dot-dashed lines), which has a slightly higher gas mass of
$M_{\rm gas}=6.03\times10^9\msun$. The results are similar to run $\rm D^{++}$.
The peaks in SFR and $M_{\rm H}$ occur slightly later, but the maximum values
are similar, and the post-starburst evolution in SFR and $M_{\rm H}$ are
also similar, even though they have very different stellar masses and gas 
fractions. 
This suggests that in isolated barred galaxies, it is 
neither the virial mass nor the total baryonic mass, but 
rather the gas mass, which is the primary factor in determining
the evolution of the galaxy. 
This is easily understood: The gas response to the non-axisymmetric 
instability is faster in gas than in stars 
\citep{berentzen_gas_2007,villa-vargas_dark_2010}. Thus at a fixed baryonic 
mass, a galaxy with higher $f_{\rm gas}$  will form a stronger bar. This 
causes the higher $f_{\rm gas}$ galaxies to have a stronger and faster 
transport of gas to the centre. The higher gas density and increased 
efficiency of radiative cooling in the central region then
favours a higher SFR.

\subsection{Resolution and Particle Number}

\begin{table*}
 \centering
 \begin{minipage}{140mm}
  \caption{Initial parameters of simulations O, D, H, OD, and HD.
           All masses are in units of $10^9\msun$}
  \begin{tabular}{@{}crrrrrrcl@{}}
  \hline
Galaxy & $M_*$ & $M_{\rm gas}$ & $M_{200}$ & 
$N_*$ & $N_{\rm gas}$ & $N_{\rm tot}$ & $f_{\rm gas}$ & colour \\
  \hline
O  &  4.0 &  1.72 &  265 &  52 501 &  22 546 &  75 047 & 0.300 
& purple solid     \\
OD &  4.0 &  1.72 &  265 & 131 146 &  56 472 & 187 618 & 0.300 
& purple dashed    \\
D  & 10.0 &  3.51 &  450 & 138 869 &  48 749 & 187 618 & 0.259 
& light green      \\
HD & 25.0 &  7.17 &  872 & 145 802 &  41 816 & 187 618 & 0.222 
& burgundy, dashed \\
H  & 25.0 &  7.17 &  872 & 364 460 & 104 583 & 469 043 & 0.222 
& burgundy, solid  \\
   \hline
\end{tabular}
\label{table:newruns}
\end{minipage}
\end{table*}

\begin{figure}
 \centering
 \includegraphics[scale=0.41]{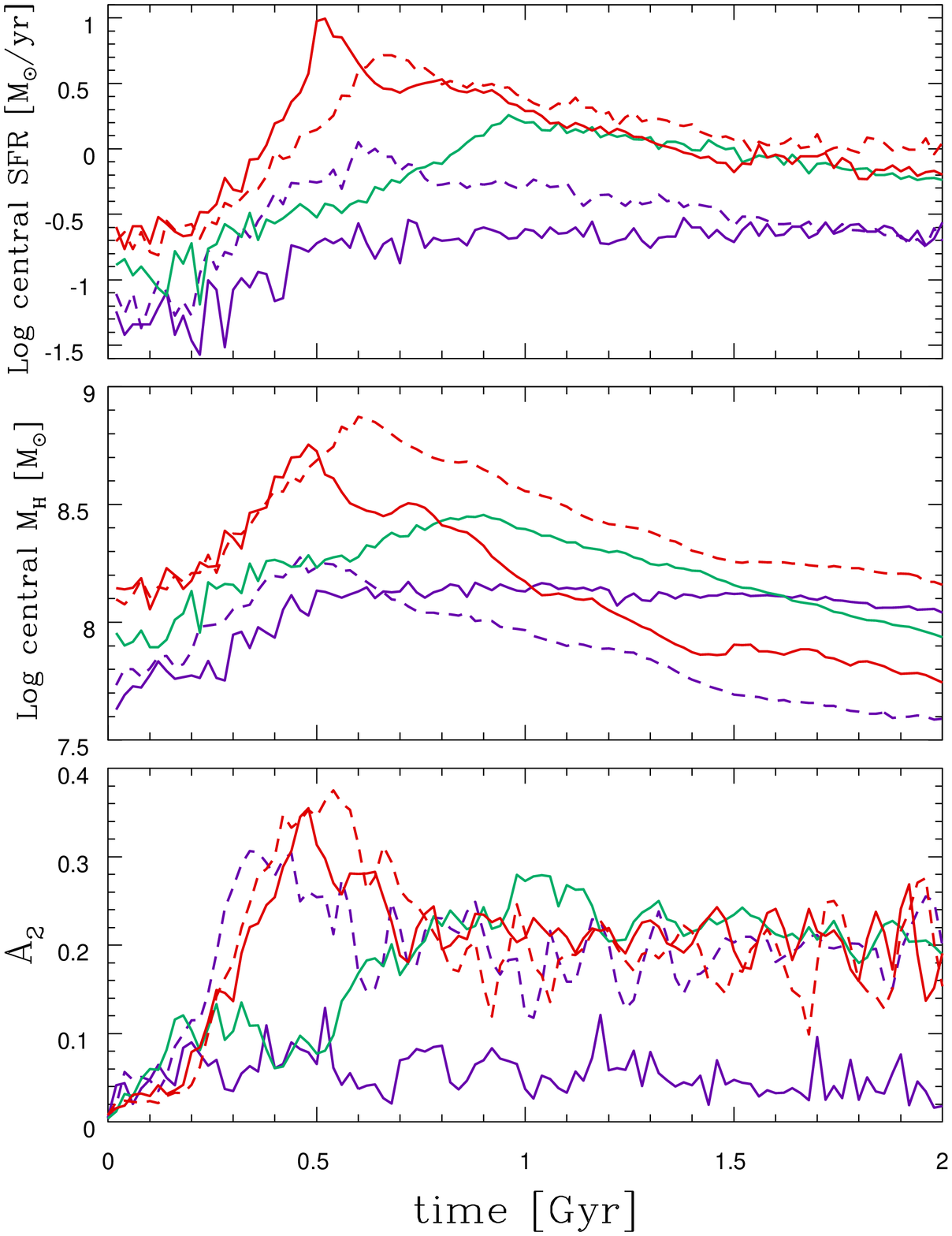} 
  \caption{Central SFR, central hydrogen mass, and bar strength vs. time
for barred galaxies H (solid burgundy), OH (dashed burgundy),
D (light green), O (solid purple), and OD (dashed purple).}
  \label{fig:D_gals}
\end{figure}


In this study, we purposely used a larger number of particles
to simulate more massive galaxies (see Table~\ref{table:initial}),
in order to maintain a fixed physical resolution across the ensemble
of simulations. In all simulations, the particle mass is of order
$72\,000\msun$.
If we had used instead the same number of particles in all simulations,
the low-mass galaxies would have a much higher physical resolution
than the high-mass ones.
This could lead to potential problems if some physical phenomena only
appear below a certain mass scale. These phenomena could be resolved in the
low-mass galaxies, but not in the high-mass ones. Also, the subgrid
model of star formation and feedback used in GCD+ 
is known to be sensitive to physical mass resolution 
\citep{kawata_numerical_2014}.
Having different physical resolutions from run to run might require an
adjustment of the subgrid parameters. 

We performed two additional simulations of barred galaxies, OD and HD,
to illustrate the effects
of varying the physical resolution. Table~\ref{table:newruns}
lists the initial parameters of these new galaxies, along with the
ones of galaxies~O, D, and~H. The seventh column shows the total number
of particles $N_{\rm tot}$. Galaxy OD
has the same mass as galaxy O, but the same total number of particles 
as galaxy D.
Similarly, galaxy HD has the same mass as galaxy H, but the same
total number of particles as galaxy D.\footnote{The number of stars
and gas particles in galaxies OD, D, and HD are different
because these galaxies have different gas fractions.}

The bottom panel of
Figure~\ref{fig:D_gals} shows the evolution of the bar strength for these
runs. Galaxy O does not form a significant bar. All other galaxies
form a bar, and though the values of $A_2$ strongly fluctuate with time,
the mean values remain around $A_2=0.2$ after $t=0.7\,\rm Gyr$.
Galaxies O has fewer particles than all the other galaxies,
which suggests that there is a minimum number
of particles required to resolve the formation of the bar.
When the bar is resolved, its strength is independent of the number of
particles used. In particular, the evolution of $A_2$
for galaxies H and HD is essentially the same, even though
galaxy H has 2.5 times more particles.
%

The upper and middle panels of Figure~\ref{fig:D_gals} show the
evolution of the central SFR and central hydrogen mass with time,
and reveal important differences between the various runs. The SFR is
significantly higher in galaxy OD than in galaxy O. This could be
explained by the fact that galaxy O simply failed to form a bar that
would drive gas inward. However, galaxy O contains three times as much
central gas as galaxy OD at the end of the simulation, indicating that
galaxy OD is much more efficient in converting gas into stars.
Comparing galaxies H and HD, we see the opposite effect: Galaxy HD has
a lower SFR than galaxy H, and is less efficient in converting central gas
into stars. This cannot be explained by a bar effect, since their
bars have the same strength.

Galaxy HD is essentially a version of galaxy H with lower physical
resolution (mass per particle: $170\,000\msun$ instead of $72\,000\msun$).
Lowering the mass resolution increases the minimum mass of gas
clumps that can form by fragmentation. These more massive clumps cannot
reach density as high as in galaxy H, and as a result star formation is
less efficient. The SFR peak is significantly lower, and is reached
later. Galaxies OD and O show the opposite effect. In this case,
the new galaxy OD has a higher physical resolution than galaxy O
(mass per particle: $30\,000\msun$ instead of $72\,000\msun$). This
leads to more fragmentation, smaller, denser gas clumps, and more
star formation. The SFR peak for galaxy OD is about $0.7\,\rm dex$ higher
than the one for galaxy O.

These comparisons show that the physical mass per particle has a major
impact on the outcome of the simulations. This vindicates our decision
to keep it uniform across the entire set of simulations.

\section{SUMMARY AND CONCLUSION}
\label{sec:conc}

We have conducted a numerical study of the star formation history
in barred and unbarred spiral galaxies, focussing on the dependence
on the total stellar mass $M_*$. We considered barred
galaxies with masses ranging from
$M_*=4\times10^9\msun$ to $M_*=2.5\times10^{10}\msun$, and for each barred 
galaxy, we simulated an unbarred galaxy of the same mass to provide a 
comparison sample.
Our main results are the following:

\begin{enumerate}

\item Barred and unbarred galaxies evolve very differently.
The bar drives a large amount of gas into the central region.
This enhances the central SFR of barred galaxies compared to
unbarred galaxies with the same total mass. The highest-mass 
barred galaxies consequently experience a starburst, with the SFR
increasing by a factor up to 30 in the central $1\rm\,kpc$
region relative to an unbarred galaxy of the same stellar mass.

\item In barred galaxies, most of the gas driven into the central region 
by the bar eventually ends up being consumed by the star formation process.
In massive barred galaxies, the strength of the starburst more than 
compensates for the fact that more gas is funnelled toward the centre,
so more massive galaxies end up with a lower central gas concentration
(see bottom panel of Fig~\ref{fig:MHF2}).
In unbarred galaxies, the lower SFR allows gas to accumulate in the
central $1\rm\,kpc$ region. As a result, unbarred galaxies are expected to
have a larger central gas concentration than barred galaxies above
a certain stellar mass $M_*$, and our results suggests
that the difference should increase with $M_*$, as more massive barred
galaxies experience stronger starbursts
(see bottom panel of Fig~\ref{fig:delta_H}).

\item Bars drive a substantial amount of gas toward the centre of barred 
galaxies, but a high efficiency enables star formation to keep up with
the build-up of gas in the central region. As a result, barred galaxies at 
late time tend to have lower central gas content and higher SFR than
unbarred galaxies of the same stellar mass.

\item We find that the initial gas mass is the main driver of the
evolution of barred galaxies. We considered galaxies with different gas 
fraction, and found that galaxies with comparable
initial gas masses had similar evolutions, even
though their virial and baryonic masses were different.

\end{enumerate}

Our simulated results broadly reproduce the observational results of 
\citet{ellison_impact_2011}. Comparing our Fig.~\ref{fig:same_fiber_young} 
with the bottom
panel of their Fig.~3, we see in the simulations and in the observations
an increase of 0.2 dex in $\rm\Delta SFR$ taking place at a total stellar
mass $\log M_*\sim9.9$. Overall, the simulated values are 
0.3 dex higher than the observed values, going from 0.3 to 0.5 instead 
of 0.0 to 0.2. But it is quite remarkable that the simulations reproduce
both the amplitude and location of the jump in $\rm\Delta SFR$.
\citet{ellison_impact_2011} suggested that star formation 
in barred galaxies is short-lived below a total stellar mass 
$\log M_*=10$, and ongoing above that mass. Our simulations suggest
an alternative explanation (see Fig.~\ref{fig:SFRF}): 
The SFR in high-mass barred galaxies
sharply increases, reaches a peak, and then slowly decays, while
the variations in SFR are less important in low-mass galaxies.
The enhancement in SFR at high-mass found in
\citet{ellison_impact_2011} is not caused by the SFR of individual 
galaxies. Instead, it is the process of averaging over galaxies that have 
different masses and 
are at different evolutionary stages, but happen to have the same central
stellar mass, that causes this enhancement.
As the top panel of Fig.~\ref{fig:SFRFm} shows, 
at large central stellar masses ($\log M_{*,\rm centre}>9.5$), we are essentially
averaging over high-mass galaxies which are in the peak of their SFR,
thus explaining the large enhancement in SFR compared to unbarred galaxies.
At lower central stellar masses, we are combining lower-mass galaxies, with a 
correspondingly lower SFR peak, with high-mass galaxies that are at an
early evolutionary stage and have not yet reached their SFR peak.

An important lesson to be learned from this study
is that one must be careful when
comparing galaxies that share one common observable property, such as
the central stellar mass. Galaxies with a same central stellar mass can have
completely different total masses, and be at very different stages
of their respective evolution. It is preferable to compare
galaxies that share at least two observable properties, 
such as gas fraction, but even that
might not be sufficient. The various curves in the top panel of
Fig.~\ref{fig:SFRFm} intersect. Hence, two galaxies with both the same central
stellar mass and the same SFR can be very different. If the
simulation of barred galaxy O was extended slightly beyond $3\,\rm Gyr$,
the violet and yellow curves would intersect. We would then have two 
barred galaxies differing by a factor of 4 in total mass, having the 
same central stellar mass and the same central SFR.
 
We have only considered isolated galaxies. In reality, galaxies can
accrete a substantial amount of matter from the intergalactic medium,
and merge with other galaxies. This could affect the dynamics of the
bar, and also affect the post-starburst evolution of barred galaxies, by
replenishing the supply of gas depleted by star formation,
although \citet{ellisonetal15} find that the 
atomic gas fraction (relative to the stellar mass) of merging galaxies is 
little changed during interactions. 
Using multi-zoom cosmological
simulations, \citet{lhuillier_mass_2012} previously studied the mass 
assembly of galaxies in a cosmological context, in order to quantify the 
respective role played by mergers and accretion. These simulations reveal
that the mass assembly history can vary wildly from one galaxy to another.
In their Section~5, they present four characteristic galaxies, and 
interestingly their galaxy b) resembles ours: the total baryonic mass
increases until redshift $z\sim1$, then remains constant for the next
$\rm 2\,Gyr$ as the gas mass decreases because of star formation
(see top right panel of their Fig.~14). 
In the last $\rm 2\,Gyr$ of its evolution, their galaxy b) essentially
evolves in isolation, just like ours (and for the same period of time as 
well). Hence, our simulations are most relevant to
low-redshift galaxies, when most of the mass assembly is completed.  
Including the effects of accretion and mergers in our simulations 
is clearly the next step in this program, and results will
be presented in forthcoming papers.

\section*{acknowledgments}
We are thankful to Laurent Drissen for useful suggestions.
This research is supported by the Canada Research Chair program and NSERC.

\label{lastpage}

\end{document}